\documentclass[11pt]{article}
\usepackage{amsmath,amssymb,color,graphics,epsfig,cite}
\usepackage{tikz}

\usepackage{amsfonts}

\newcommand{\be}{\begin{equation}}
\newcommand{\ee}{\end{equation}}
\newcommand{\bea}{\setlength\arraycolsep{2pt} \begin{eqnarray}}
\newcommand{\eea}{\end{eqnarray}}
\newcommand{\nn}{\nonumber}

\def\cramp{\medmuskip = 2mu plus 1mu minus 2mu}

\def\ft#1#2{{\textstyle{\frac{\scriptstyle #1}{\scriptstyle #2} } }}
\def\fft#1#2{{\frac{#1}{#2}}}

\def\0{{\sst{(0)}}}
\def\1{{\sst{(1)}}}
\def\2{{\sst{(2)}}}
\def\3{{\sst{(3)}}}
\def\4{{\sst{(4)}}}
\def\5{{\sst{(5)}}}
\def\6{{\sst{(6)}}}
\def\7{{\sst{(7)}}}
\def\8{{\sst{(8)}}}
\def\sst#1{{\scriptscriptstyle #1}}

\begin{document}

\begin{center}
{\Large {\bf Ricci-flat Seeds, Trombone Rescaling \\ and Magnetizing Kerr Black Holes}}

\vspace{20pt}

Liang Ma and H. L\"u

\vspace{10pt}

{\it Center for Joint Quantum Studies, Department of Physics,\\
School of Science, Tianjin University, Tianjin 300350, China }

\vspace{40pt}

\underline{ABSTRACT}
\end{center}

We present a new family of Ricci-flat rotating metrics, which we refer to as Solution II, obtained by demagnetizing the genuine Kerr-Bertotti-Robinson (G-KBR) black hole. Together with Solution I, which was previously obtained by demagnetizing the Kerr-Bertotti-Robinson (KBR) black hole, this gives two families of Ricci-flat deformations of the Kerr black hole, each labeled by $B$. We study the global structures, thermodynamics, various limits, and the local connections among these limits. We also study the remagnetizations in both Einstein-Maxwell and Kaluza-Klein theories, and find that, to yield a physically acceptable solution, the Ernst transformation should be augmented by an appropriate trombone rescaling.

\vfill {\footnotesize maliang0@tju.edu.cn\ \ \ mrhonglu@gmail.com}

\thispagestyle{empty}
\pagebreak

\tableofcontents
\addtocontents{toc}{\protect\setcounter{tocdepth}{2}}

\newpage

\section{Introduction}

Einstein-Maxwell theory incorporates the interplay between two long-range forces in nature: gravity and electromagnetism. The static Reissner-Nordstr\"om \cite{Reissner:1916cle, Nordstrom:1918lpd} and rotating Kerr-Newman \cite{Newman:1965my} black holes are well-known examples of coupled gravitational and Maxwell-field configurations in which these interactions are balanced. Beyond these charged black holes, Einstein-Maxwell theory admits a further class of configurations in which black holes are immersed in external magnetic fields. Since astrophysical objects in the Universe are often exposed to strong magnetic fields, there has been a continuing effort in the construction and study of such exact solutions.

In general, black holes immersed in external magnetic fields are constructed by applying an Ernst transformation to a Ricci-flat black hole seed solution. The well-known Schwarzschild-Melvin and Kerr-Melvin black holes are prominent examples of this approach \cite{Bonnor:1954tis,Melvin:1963qx,Ernst:1976mzr,Ernst:1976bsr}. More recently, the KBR black hole carrying electric charges and the uncharged G-KBR black hole have been constructed \cite{Podolsky:2025tle,Ovcharenko:2025cpm,Ovcharenko:2026tos,Kubiznak:2026uro}.  Unlike the Melvin universe, which describes a magnetic universe with a magnetic field aligned along the \(z\)-axis, the Bertotti-Robinson universe in which these black holes are embedded is an electrovacuum characterized by a uniform magnetic field. Moreover, neither the KBR nor the G-KBR was generated through the conventional Ernst transformation; instead, both were obtained by directly solving the Einstein-Maxwell equations. The properties of these black holes have since been the subject of extensive investigation \cite{Astorino:2025lih,Astorino:2026okd,Ovcharenko:2026byw, Barrientos:2026shy, Hu:2026slp,Herdeiro:2026jem,DiPinto:2026rvp,Lyu:2026pqh,Huang:2026qzj, Li:2026gvm,DiPinto:2026ynu,Astorino:2026nhd,Furugori:2026qbb,Ma:2026otg,Zhou:2026xrc}.

Einstein-Maxwell theory is very special in that, when reduced on a circle via the Kaluza-Klein (KK) ansatz, the resulting three-dimensional theory is a nonlinear sigma model, described by a scalar coset of $SU(2,1)/(SU(2)\times U(1))$. The Ernst transformation in this sigma-model description becomes a specific $SU(2,1)$ transformation. In particular, the Kerr-Melvin solution can be obtained via an $SU(2,1)$ transformation \cite{Gibbons:2013yq}
\be
U = \begin{pmatrix}
1 & 0 &\; 0 \\
\displaystyle \frac{\mathcal{B}}{\sqrt{2}} & 1 &\; 0 \\
\displaystyle \frac{\mathcal{B}^2}{4} & \displaystyle \frac{\mathcal{B}}{\sqrt{2}} &\; 1
\end{pmatrix}.\label{Magnetizing U}
\ee
Although both the KBR and G-KBR black holes reduce to the Kerr solution when the external magnetic field parameter \(B\) vanishes, neither can be obtained by applying the above \(SU(2,1)\) transformation to the Kerr black hole, since it gives the Kerr-Melvin solution. This raises a natural question: can these magnetized KBR or G-KBR black holes be generated from an underlying Ricci-flat black hole geometry? Since the transformation above acting on the Kerr metric gives rise to the Kerr-Melvin black hole, there are only two possible answers other than ``no.'' One is that the Ricci-flat seed remains the Kerr metric, but the $SU(2,1)$ transformation should be modified. The other is that there exists a new Ricci-flat seed. As shown in \cite{Ma:2026otg}, demagnetizing the KBR black hole using \eqref{Magnetizing U} yields a novel Ricci-flat solution characterized by an additional integration constant \(B\), in addition to the mass and rotation parameters \((\mu,\, a)\). This solution can be interpreted as a \(B\)-deformed Kerr black hole, which we refer to as Solution I. It is therefore natural to ask whether the G-KBR black hole likewise admits a corresponding  Ricci-flat counterpart. One key finding of this paper is that such a Ricci-flat seed indeed exists, and we refer to this metric as Solution II.

Both Ricci-flat Solution I and Solution II are one-parameter deformations of the Kerr metric, with the parameter denoted by B in both cases. Their local and global structures are very complicated and we analyse them in some detail together for comparison. These Ricci-flat metrics can be used as seeds for magnetization not only in Einstein-Maxwell theory but also in KK theory, both of which are special cases of the more general Einstein-Maxwell-Dilaton (EMD) theory
\be
{\cal L} = \sqrt{-g} (R - \ft12 (\partial\phi)^2 - \ft14 e^{a\phi} F^2)\,,\qquad F=dA\,.
\label{emd}
\ee
KK theory corresponds to $a=\sqrt3$ and Einstein-Maxwell theory can be embedded into EMD theory with $a=0$. There exists an $SL(2,\mathbb R)$ solution-generating technique for magnetizing any static Ricci-flat metric \cite{Yazadjiev:2005gs}. (See \cite{Ivashchuk:2013jja,Bolokhov:2017rnf,Ma:2026rro,Hong:2026hsa} for a variety of examples of electrovacua or fluxbranes in EMD theories or their generalizations.) However, for rotating solutions such as the Kerr metric, or the $B$-deformed Kerr metrics, such techniques exist only in Einstein-Maxwell theory or in KK theory. (See \cite{Ma:2026rro,Hong:2026hsa}.)

Subtleties emerge if we use the solution-generating techniques to magnetize Solution I and Solution II. In the case of magnetizing the Kerr metric, we obtain the Kerr-Melvin solution where the magnetic field parameter is simply the parameter ${\cal B}$ of the Ernst transformation. In order to obtain KBR or G-KBR solutions from Solution I and Solution II, respectively, the transformation parameter ${\cal B}$ must be a special function of the Ricci-flat metric parameter $B$, i.e., ${\cal B}={\cal B}(B)$. After setting this relation, we then find that the resulting metric may not have a smooth $B\rightarrow 0$ limit, nor a smooth $a\rightarrow 0$ limit, where $a$ is the rotation parameter of the Kerr metric. In order to obtain a smooth limit, another trombone transformation must be applied, namely
\be
g_{\mu\nu} \rightarrow \Lambda^2\, g_{\mu\nu}\,,\qquad A_\mu \rightarrow \Lambda\, A_\mu\,,
\label{trombone}
\ee
where the conformal factor $\Lambda^2$ is a constant, which rescales the whole action; it is therefore a symmetry of the equations of motion. Thus, in these more complicated magnetization procedures, the Ernst transformation must also include the trombone transformation.

The paper is organized as follows. In Section \ref{sec:Class I}, we focus on the global structure of Solution I, obtained in \cite{Ma:2026otg}. In the standard $(r,x=\cos\theta)$ coordinates of the Kerr metric, the metric has two ``$\pm$'' branches. These two branches can be unified in the new $(r,\Omega)$ coordinates for nonzero $B$, but they become distinct when $B=0$. The global structure of the $+$ branch was studied in detail in \cite{Ma:2026otg}, and we present a global analysis of the ``$-$'' branch. In Section \ref{sec:Class II-local}, we present our key finding of this paper, the local metric of Solution II. We present strong evidence that the metric is locally inequivalent to Solution I. In Section \ref{sec:Class II-global}, we analyse the global structure of Solution II, which also has $\pm$ branches. In Section \ref{sec:B=0}, we take the $B=0$ limit, in which case both ``$+$'' branches yield the same Kerr metric, but the ``$-$'' branches give rise to two new rotating generalizations of the Schwarzschild-Levi-Civita metric, and we show it is locally inequivalent to the metric already constructed in the literature. In Section \ref{sec:magnetization}, we study the magnetization and present the required trombone transformation. We conclude the paper in Section \ref{sec:conclusion}, where we present a figure illustrating Solutions I and II, their various limits and connections.

\section{Solution I: local metric and global structure}
\label{sec:Class I}

\subsection{The local metric}

Before presenting our new findings, we first review our previously constructed Ricci-flat metric, since there are additional global properties worth further study. This material will be useful for comparison with the new solution presented in Sections \ref{sec:Class II-local} and \ref{sec:Class II-global}. We refer to the Ricci-flat metric constructed in \cite{Ma:2026otg} as Solution I, given by
\bea
ds_{\rm I}^2&=&\frac{P r^4 (1-x^2)-a^2 Q x^4}{\mathbb{L}_{1}\Omega^4}\Big[
d\varphi+\frac{a\mathbb{M}_{1}}{P r^4 (1-x^2)-a^2 Q x^4}dt
\Big]^2\cr
&&-\frac{(1-x^2)PQ\mathbb{L}_{1}}{P r^4 (1-x^2)-a^2 Q x^4}dt^2+\mathbb{L}_{1}\Big(\frac{dr^2}{Q}+\frac{dx^2}{P(1-x^2)}\Big)\,.\label{Rotating Ricci flat class 1}
\eea
Note that the vielbein structure presented differs from the one given in \cite{Ma:2026otg}. The two functions \(\mathbb{L}_{1}\) and \(\mathbb{M}_{1}\) are given by
\bea
\mathbb{L}_1&=&\frac{1}{4\Omega^2}\big[r^2(1+\Omega-r\partial_r\Omega)^2+a^2x^2(1+\Omega-x\partial_x\Omega)^2\big]\,,\cr
\mathbb{M}_1&=&-\frac{1}{2}\Big[
Pr^2(1-x^2)(1+\Omega-x\partial_x\Omega)+Qx^2(1+\Omega-r\partial_r\Omega)
\Big]\,,\label{M1L1}
\eea
together with
\bea
\Sigma&=&r^2+a^2x^2\,,\qquad
P=1+B^2\Big(\mu^2\frac{I_2}{I_1^2}-a^2\Big)x^2\,,\qquad Q=(1+B^2r^2)\Delta\,,\cr
\Omega^2&=&(1+B^2r^2)-B^2\Delta x^2\,,\qquad \Delta=\Big(1-B^2\mu^2\frac{I_2}{I_1^2}\Big)r^2-2\mu\frac{I_2}{I_1}r+a^2\,,\cr
I_1&=&1-\frac{1}{2}B^2a^2\,,\qquad I_2=1-B^2a^2\,.\label{4D solution 2 class 1}
\eea
The metric \eqref{Rotating Ricci flat class 1} is of cohomogeneity two, depending on two coordinates $(r, x)$. The remaining coordinates $(t, \varphi)$ correspond to two Killing directions. In terms of the $(r, x)$ coordinates, the metric has two branches, associated with taking the $\pm$ roots of $\Omega^2$, namely
\be
\Omega_{\pm}=\pm \sqrt{\Omega^2} = \pm\sqrt{(1+B^2r^2)-B^2\Delta x^2}\,.
\ee
These two branches arise from taking the $(r,x)$ coordinate choice. If we use $(r, \Omega)$ as coordinates instead, the coordinate $\Omega$ naturally takes all real values, both positive and negative, since the coordinate $x$ appears in the original metric via $x^2$.
However, the metric in $(r,\Omega)$ coordinates has an unappealing feature: the off-diagonal term $g_{r\Omega}$ does not vanish. A much more serious defect is that in the limit $B=0$, we have $\Omega=1$. Thus, it is advantageous in general to use the $(r,x)$ coordinates rather than $(r,\Omega)$ coordinates, so we should analyse the metric separately in the two $\Omega_\pm$ branches. For the $\Omega_+$ branch, we denote the metric simply as
\be
ds_{\rm I,+}^2 = ds_{\rm I}^2(\Omega_+)\,.\label{classIplus}
\ee
This metric was first constructed and analysed in \cite{Ma:2026otg}, and it has a smooth $B\rightarrow 0$ limit, reducing to the Kerr black hole. The metric was therefore referred to as the $B$-deformed Kerr black hole. Many properties of the metric \eqref{classIplus} were analysed in \cite{Ma:2026otg}, where it was shown to describe a black hole whose thermodynamics satisfies the first law.

In this section, we shall focus largely on the properties of the $\Omega_-$ branch. The metric \eqref{Rotating Ricci flat class 1}, obtained by simply taking $\Omega=\Omega_-$, has no smooth $B\rightarrow 0$ limit. In order for the metric to have a smooth $B\rightarrow 0$ limit, and then a smooth static limit $a\rightarrow 0$, we must take a trombone rescaling discussed in the introduction. Specifically, we have
\be
ds_{\rm I,-}^2=\frac{16}{B^4}ds^2_{\rm I}(\Omega_-)\,,\qquad \hbox{followed by}\qquad  \varphi\quad \rightarrow\quad \varphi'=\frac{B^4}{16}\Big(\varphi-\frac{4a}{B^2}t\Big)\,.
\ee
As discussed, modulo the trombone rescaling, the two metrics $ds_{\rm I, \pm}^2$ can be unified into the metric $ds_{\rm I}^2$ under $(r, \Omega)$ coordinates for a nonzero parameter $B$. When $B=0$, $\Omega=\pm 1$ no longer corresponds to a coordinate. Thus, the metrics $ds_{\rm I,+}^2$ and $ds_{\rm I, -}^2$ become locally distinct when $B=0$. We shall further discuss this $B=0$ limit in Section \ref{sec:B=0}.

\subsection{Static limit}

It is instructive to study the static $a=0$ limit, in which case the metric in either of the two branches becomes significantly simpler. We have
\bea
&&ds_+^2 \equiv ds_{\rm I, +}^2\Big|_{a\rightarrow 0}\nn\\
&&= \frac{\big(1+B^2\mu rx^2+\Omega_+\big)^2}{4\Omega_+^4}\Big[
-\frac{Q}{r^2}dt^2+\frac{r^2}{Q}dr^2+\frac{r^2}{P}\frac{dx^2}{1-x^2}
\Big]+\frac{4r^2(1-x^2)P}{\big(1+B^2\mu rx^2+\Omega_+\big)^2}d\varphi^2\,,\nn\\
&&ds_{-}^2 \equiv ds_{\rm I,-}^2\Big|_{a\rightarrow 0}\label{two static Ricci flat}\\
&&=\frac{4\big(1+B^2\mu rx^2+\Omega_-\big)^2}{B^4\Omega_-^4}\Big[
-\frac{Q}{r^2}dt^2+\frac{r^2}{Q}dr^2+\frac{r^2}{P}\frac{dx^2}{1-x^2}
\Big]+\frac{B^4r^2(1-x^2)P}{4\big(1+B^2\mu rx^2+\Omega_-\big)^2}d\varphi^2\,,\nn
\eea
where
\bea
P&=&1+B^2 \mu ^2 x^2\,,\qquad \Omega_\pm=\pm \sqrt{1+B^2 \Big(\mu  r  \big(B^2 \mu  r+2\big)x^2+r^2 \big(1-x^2\big)\Big)}\,,\cr
Q&=&r^2\big(1+B^2 r^2\big) \Big(1-B^2 \mu ^2-\frac{2 \mu }{r}\Big)\,.
\eea
The metric $ds_+^2$ was studied in \cite{Astorino:2025lih,Astorino:2026okd} (see also \cite{Ma:2026otg}), and for $B\ne 0$, the metric $ds_-^2$ is the other branch, up to trombone rescaling. The $ds_-^2$ metric was studied in \cite{Barrientos:2026shy}. However, in the $B=0$ limit, the metrics $ds_+^2$ and $ds_-^2$ reduce to the Schwarzschild metric and the Schwarzschild-Levi-Civita (SLC) metric \cite{Stephani:2003tm}, respectively, i.e.
\bea
&&ds_+^2\Big|_{B=0} = ds_{\rm Sch}^2 = -(1-\frac{2\mu}{r})dt^2+\frac{dr^2}{1-\frac{2\mu}{r}}+r^2\frac{dx^2}{1-x^2}+r^2 (1-x^2)\, d\varphi^2\,,\label{Schmetric}\\
&&ds_-^2\Big|_{B=0} = ds_{\rm SLC}^2 = r^4(1-x^2)^2\Big[-(1-\frac{2\mu}{r})dt^2+\frac{dr^2}{1-\frac{2\mu}{r}}+r^2\frac{dx^2}{1-x^2}
\Big]+\frac{d\varphi^2}{r^2(1-x^2)}\,.\label{Sch-LC}
\eea
Thus, the $ds_+^2$ and $ds_-^2$ metrics can be viewed as the $B$-deformed Schwarzschild and SLC metrics, respectively; they can themselves be unified as two branches of one local solution in $(r,\Omega)$ coordinates. However, when $B=0$, the resulting Schwarzschild and SLC metric are locally distinct.

It is well understood that the Schwarzschild metric has a curvature singularity at $r=0$. The locations $x=\pm 1$ are the north and south poles of the foliating $S^2$ surfaces. The situation for the SLC metric is quite different. The square of the Riemann tensor is
\be
\mathrm{Riem}^2=\frac{48 \big[4 r^2-6 r (3-4 x^2+x^4) \mu +21 (1-x^2)^2 \mu ^2\big]}{r^{14} (1-x^2)^6}\,.
\ee
Thus, not only is the $r=0$ a curvature singularity, but the would-be north and south poles $x=\pm 1$ are also singularities. The situation is analogous for the $B$-deformed metrics. The metric $ds_+^2$ has a curvature singularity at $r=0$ when $\mu\ne 0$. The metric $ds_-^2$, on the other hand, has curvature singularities at $r=0$ as well as at $x=\pm 1$, as we can see from the leading divergence term of the Riemann tensor squared:
\be
\mathrm{Riem}^2\sim\frac{192(1+B^2\mu r)^{12}}{r^{12}(1+B^2\mu^2)^2}\frac{1}{(1-x^2)^6}\,.\label{Sch B Riem2}
\ee

To understand the two branches of the static metrics $ds_\pm^2$ better, we consider a further $\mu=0$ limit while keeping $B$ nonzero. In real $(r,x)$ coordinates, the $ds_+^2$ metric has no singularity, but the $ds_-^2$ metric has a singularity at $\Omega_-=-1$. It was observed \cite{Astorino:2026okd} that the metric $ds_+^2$ with $\mu=0$ can be locally mapped to a Ricci-flat metric in the Kundt class, namely
\be
ds^2_{\rm K}=p^2\Big[-(1+B^2q^2)dt^2+\frac{dq^2}{1+B^2q^2}\Big]
+\frac{p}{B^2(1-p)}dp^2+\frac{4(1-p)}{B^2p}d\varphi^2\,.\label{Kundt class}
\ee
The corresponding coordinate transformation from $(r,x)$ to $(q,p)$ is
\be
r=\frac{\sqrt{4p-4p^2+B^2q^2}}{B(2p-1)}\,,\qquad x=\frac{Bq}{\sqrt{4p-4p^2+B^2q^2}}\,,\label{rxtopq}
\ee
The quantity $\Omega^2$ under this coordinate transformation is
\be
\Omega^2 = 1 + B^2 r^2 (1-x^2) = \fft{1}{(2p-1)^2}\,.
\ee
Thus, we see that the $\Omega_+>0$ branch requires $r>0$ and hence $p>\frac 12$. (The $r<0$ and hence $p<\frac 12$ region corresponds to the Kundt metric with $p\rightarrow 1-p$.)

The $\Omega_-$ branch requires a trombone rescaling first to become $(B^4/16)\, ds_-^2$, which can then be mapped to the Kundt metric \eqref{Kundt class} with the same $(r,x)$  coordinate transformation provided that $r<0$ and hence $p<\frac 12$. Note that a further $\varphi$ transformation $\varphi\rightarrow B^4\varphi/16$ is also needed.  (Analogously, the $r>0$ and hence $p>\frac 12$ region corresponds to the Kundt metric with $p\rightarrow 1-p$.) It is now clear that the joining of the two branches in the $(p,q)$ Kundt coordinates at $p=1/2$ turns into a splitting into $\Omega_\pm =\pm \infty$ in the $(r, \Omega)$ coordinates; while the joining at $\Omega=0$ in the $(r, \Omega)$ coordinates becomes a splitting into $p=\pm \infty$ in the $(p,q)$ coordinates. The singularity $p=0$ of the Kundt metric can only be reached by the $\Omega_-$ branch, associated with $x=\pm 1$. The regular north and south poles $x=\pm 1$ of the $\Omega_+$ branch are mapped to $p=1$, which is a coordinate singularity, but is regular in curvature tensor. In the $B=0$ limit, the coordinate transformation \eqref{rxtopq} collapses, and hence the connection between $ds_{+}^2$ and $ds_{\rm -}^2$ is severed.

\subsection{$ds_{\rm I,-}^2$: global analysis and black hole thermodynamics}

We now return to the rotating metric $ds_{\rm I}^2$ in \eqref{Rotating Ricci flat class 1} with nonzero parameter $a$. The global structure of the $ds_{\rm I, +}^2$ branch was analysed in our previous paper \cite{Ma:2026otg}, where we showed that it describes a black hole and obtained thermodynamic quantities satisfying the first law. In this subsection, we therefore focus on the analysis of the $ds_{\rm I,-}^2$ branch. As discussed, its static limit $ds_{-}^2$ describes the $B$-deformed SLC metric, where $x=\pm 1$ are curvature singularities. Thus, the corresponding topology of a two-dimensional constant-$(t,r)$ slice is not a 2-sphere but a singular space with singularities at $x=\pm 1$. Interestingly, these two singularities are resolved when we turn on the rotation parameter $a$ in $ds_{\rm I,-}^2$. By resolution, we mean that the $x=\pm 1$ become coordinate singularities rather than curvature singularities in the metric $ds_{\rm I,-}^2$. In other words, a spacelike Killing vector becomes degenerate at $x=\pm 1$. To see this explicitly, it is advantageous to redefine the two Killing coordinates such that
\be
\varphi\quad \rightarrow\quad \varphi'=\frac{\varphi}{P_{0,\pm}}\,,\qquad t\quad \rightarrow \quad t'=t+\alpha_\pm\frac{\varphi}{P_{0,\pm}}\,.\label{remove CTC and conical}
\ee
For \(ds_{\rm I,+}^2\), the quantities \(\{\alpha_+,P_{0,+}\}\) were discussed and given in \cite{Ma:2026otg}. For \(ds_{\rm I,-}^2\), we find that \(\{\alpha_-,P_{0,-}\}\) are given by
\be
\alpha_-=-\frac{aB^4}{4(1-\sqrt{I_2})^2}\,,\qquad P_{0,-}=\frac{B^4I_2(I_1^2+B^2\mu^2)}{4I_1^2(1-\sqrt{I_2})^2}\,.\label{remove CTC and conical 2}
\ee
Consequently, the Killing vector that becomes degenerate at $x=\pm 1$ is simply $\partial/\partial \varphi$, and it generates a period of $2\pi$. Thus, the constant-$(r,t)$ slice is again a two-dimensional space with 2-sphere topology in the coordinates $(x,\varphi)$. However, in the $a=0$ limit, the ratio \(\alpha_-/P_{0,-}\) remains finite, whereas \(P_{0,-}\) diverges. This implies that the regularity at \(x=\pm1\) is incompatible with the \(2\pi\) periodicity of \(\varphi\), and the metric becomes singular at $x=\pm 1$ for $a=0$, as discussed earlier. It is rather unusual that a spacetime curvature singularity is resolved by simply introducing rotation.

However, there could be a price to pay for resolving the $x=\pm 1$ naked singularities of the SLC metric by turning on rotation, which may give rise to close timelike curves (CTCs). If there are CTCs owing to rotation, then the issue is whether the CTCs are naked or shielded by the event horizon. To determine this, we need to examine the sign of $g_{\varphi\varphi}$ obtained after the coordinate redefinition \eqref{remove CTC and conical}. We find that it is now
\bea
g_{\varphi\varphi}&=&\frac{4I_1^4}{B^4I_2^2(I_1^2+B^2\mu^2)^2}\,\frac{1}{r^4(1-x^2)P-a^2x^4Q}
\nn\\
&&\bigg\{
\frac{1}{\mathbb{L}_1 \Omega ^4}\Big[(1-\sqrt{I_2}) \big(P r^4 (1-x^2)-a^2 Q x^4\big)-2 a^2 \mathbb{M}_1\Big]^2-4 a^2(1-x^2) P Q  \mathbb{L}_1
   \bigg\}.
\eea
The metric would have CTCs if $g_{\varphi\varphi}<0$. A physically acceptable metric should have no naked CTCs outside the event horizon; we shall address this issue first.

To discuss the event horizon, we note that besides the two degenerate surfaces at $x=\pm 1$, the metric also has degenerate surfaces at the two roots of the quadratic function $\Delta(r)$. However, we must impose that \cite{Ma:2026otg}
\be
\mu<\mu_{\max} \equiv \fft{I_1}{B\sqrt{I_2}}\,,
\ee
so that the coefficient of $r^2$ in $\Delta(r)$ is positive. This requirement ensures that $r$ is a non-compact coordinate extending to $r\rightarrow \infty$, since $\Delta(r)$ is positive for large $r$.  The quadratic function $\Delta(r)$ has two roots, namely \cite{Ma:2026otg}
\be
r_\pm = \frac{I_1 (I_2 \mu \pm \sqrt{I_2\mu^2 - a^2 I_1^2})}{B^2I_2(\mu_{\rm max}^2-\mu^2)}\,.
\ee
The reality condition requires that we impose
\be
\mu \ge \mu_{\rm ext} \equiv \fft{a I_1}{\sqrt{I_2}}\,,\qquad\rightarrow\qquad aB<1
\quad\hbox{for}\quad \mu_{\rm max}> \mu_{\rm ext}\,.
\ee
When $\mu_{\rm max}>\mu>\mu_{\rm ext}$, there are in general two positive roots, and we refer to the larger root $r_+$ as $r_h$, the radial location of the event horizon. The two roots $r_\pm$ coalesce when $\mu=\mu_{\rm ext}$, which corresponds to the extremal case.

Having established the event horizon, we are now in a position to return to the issue of CTCs. The region on and outside the horizon is given by $r\ge r_h$. We find that $g_{\varphi\varphi}$ is always positive in the vicinity of the event horizon $r_h$. Depending on the choice of parameters $(B, \mu, a)$, $g_{\varphi\varphi}$ can stay positive for all $r\ge r_h$. However, we find that for other choices, velocity of light surfaces develop at $x^2=x_*^2<1$ for sufficiently large $r$, where $g_{\varphi\varphi}=0$ and $g_{\varphi\varphi}<0$ for $x\in (-x_*, x_*)$. To determine the restrictions on $(B,\mu, a)$, we note that if naked CTCs emerge, then the quantity $g_{\varphi\varphi}$ at $x=0$ and $r\rightarrow \infty$ is always the most negative. We therefore derive a constraint for the absence of naked CTCs, namely
\be
\mu > \mu_{\rm ctc} \equiv \fft{I_1 \sqrt{5\sqrt{I_2}-3}}{B \sqrt{I_2(1+\sqrt{I_2})}}\,.
\ee
It is of interest to note that we always have
\be
\mu_{\rm max}>\mu_{\rm ctc}\,,\qquad \mu_{\rm max}> \mu_{\rm ext}\,.
\ee
However, the relative size of $\mu_{\rm ctc}$ and $\mu_{\rm ext}$ depends on the specific values of $(B,\mu,a)$. To classify the spectrum, we define a purely numerical quantity $\eta$:
\be
\eta = \ft13 (z + 7 z^{-1} - 4) \sim 0.658967\,,\qquad z=(44 + 3\sqrt{177})^{1/3}\,.
\ee
For $\eta<aB <1$, we always have $\mu_{\rm ext}>\mu_{\rm ctc}$, and all black holes have no naked CTCs. (Note that $\mu_{\rm ctc}^2\le 0$ for $aB\ge 4/5$, and hence the condition $\mu_{\rm ext}^2> \mu _{\rm ctc}^2$ is trivially satisfied.) For $0<aB<\eta$, we have $\mu_{\rm ext}<\mu_{\rm ctc}$. In this case, black holes with $\mu>\mu_{\rm ctc}$ have no naked CTCs, but black holes with $\mu\le \mu_{\rm ctc}$ develop naked CTCs. Thus it can be said that the majority of black holes described by $ds_{\rm I,-}^2$ have no naked CTCs. By comparison, the metric $ds_{\rm I,+}^2$ has no naked CTCs at all, as shown in \cite{Ma:2026otg}.

We now turn to the discussion of black hole thermodynamics. The metric $ds_{\rm I, -}^2$ suffers from a number of undesirable features, including the lack of an asymptotic Minkowski region so that we cannot define a proper scaling of the time coordinate and hence of the mass or temperature. However, as discussed above, it is quite nontrivial that the naked CTCs can be avoided for a large class of black holes in a suitable parameter space. Following the same technique as for \(ds_{\rm I,+}^2\) in \cite{Ma:2026otg}, we compute the corresponding thermodynamic quantities. Starting from the coordinates \eqref{remove CTC and conical}, where both CTCs and conical singularities have been removed, we further perform the most general coordinate transformation among the Killing coordinates
\bea
t\quad \rightarrow \quad t=\lambda_1t\,,\qquad \varphi\quad \rightarrow\quad \varphi'=\varphi+\lambda_2Bt\,.\label{coord first law 1}
\eea
This transformation does not affect $g_{\varphi\varphi}$. Using standard methods, we can now determine the temperature, horizon angular velocity, and entropy, all of which depend on the horizon location $r_h$.
\bea
T&=&\lambda _1\frac{(1-\sqrt{I_2}) (1+B^2 r_h^2) (I_2 r_h^2-a^2) \big(1+\sqrt{1+B^2 r_h^2}\big)}{4\pi B^2
   r_h^3 \big(\sqrt{I_2} r_h^2+a^2 \sqrt{1+B^2 r_h^2}\big)}\,,\cr
\omega_H&=&B\lambda_2-\lambda_1\frac{a \sqrt{1+B^2 r_h^2} \big(\sqrt{I_2}+\sqrt{1+B^2 r_h^2}\big) \big(I_1 \sqrt{1+B^2
   r_h^2}-\sqrt{I_2}\big)}{(1-\sqrt{I_2}) (r_h^2+a^2) \big(1+\sqrt{1+B^2 r_h^2}\big)}\,,\cr
S&=&\frac{4 \pi  r_h^4 a^2 \big(\sqrt{I_2}-\sqrt{1+B^2 r_h^2}\big)}{(1+\sqrt{I_2}) (1+B^2 r_h^2)^{3/2}
   \big(1+\sqrt{1+B^2 r_h^2}\big) \big(\sqrt{I_2}-I_1 \sqrt{1+B^2 r_h^2}\big)}\,.
\eea
Then, following the generalized Komar integration approach introduced in \cite{Liu:2022wku}, we compute the conserved charge
\be
\mathbf{Q}=-{*d\xi}=-V(r,x) dx\wedge d\varphi + U(r,x) dr\wedge d\varphi\,,\qquad
\hbox{satisfying}\qquad d\mathbf{Q}=0\,,
\ee
associated with the Killing vector $\xi = c_1 \partial_t - c_2 \partial_\varphi$, i.e.,
\bea
c_1M+2c_2J=\frac{1}{8\pi}\int \mathbf{Q}=\frac{\int d\varphi}{8\pi}\Big(
\int_{-1}^{+1}V(r,x)dx+\int_{r_+}^rU(r',x)\Big|_{x=-1}^{x=+1}dr'
\Big)\,.\label{Komar Ricci flat}
\eea
The expressions of $V(r,x)$ and $U(r,x)$ are very complicated and we shall not present them here. The mass and angular momentum are given by
\bea
J&=&\frac{4\mu a I_1^3(1-\sqrt{I_2})^2  }{B^4 I_2^{3/2} (I_1^2+B^2 \mu ^2)^2}\,,\qquad
M=2\lambda_2BJ-\lambda_1\frac{4\mu I_1 (1-\sqrt{I_2})^2  }{B^4\sqrt{I_2} (I_1^2+B^2 \mu ^2)}\,.
\eea
It is worth mentioning that two quantities, the entropy and angular momentum, are independent of the parameters $\lambda_1$ and $\lambda_2$. The thermodynamic quantities we obtained above automatically satisfy the Smarr relation for any $\lambda_1$ and $\lambda_2$:
\be
M=2TS+2\omega_HJ\,.
\ee
This is not surprising, since the Smarr relation follows directly from the conservation of the Komar integration, derived from $d\mathbf{Q}=0$.

We now require these thermodynamic quantities to satisfy both the first law and the Christodoulou-Ruffini (CR) relation \cite{Christodoulou:1971pcn}, namely
\be
\delta M=T\delta S+\omega_H\delta J\,,\qquad M^2=\frac{S}{4\pi}+\frac{\pi J^2}{S}\,.
\label{firstlawCR}
\ee
We find that these two conditions can be satisfied, provided that the two parameters \(\lambda_{1,2}\) are given by
\bea
\lambda_1&=&\frac{B^2 \big[2 \alpha ^2-2 \alpha  (1+\alpha ) \beta +(1+\alpha -\alpha ^2+\alpha ^3) \beta ^2\big]}{2 (1-\alpha)^{3/2}
   \sqrt{\beta } \sqrt{2 \alpha -2 \alpha  (1+\alpha ) \beta +(1-\alpha +\alpha ^2+\alpha ^3) \beta ^2} \sqrt{(1+\alpha
   ^2) \beta -2 \alpha }}\,,\cr
\lambda_2&=&\frac{B^2 \sqrt{1+\alpha } \sqrt{\beta } \big[\alpha -2 \alpha \beta + (1-\alpha+\alpha^2)\beta^2\big] \sqrt{(1+\alpha ^2) \beta
   -2 \alpha }}{2 (1-\alpha )^2 (\beta ^2-1) \sqrt{2 \alpha -2 \alpha  (1+\alpha ) \beta +(1-\alpha +\alpha ^2+\alpha ^3)
   \beta ^2}}\,,
\eea
where for convenience, we introduced two dimensionless parameters, \(\alpha\) and \(\beta\), given by
\be
\alpha=\sqrt{1-a^2B^2}\,,\qquad \beta=\sqrt{1+B^2r_h^2}\,.
\ee
When \(\alpha\beta=1\), corresponding to $\mu=\mu_{\rm ext}$, the black hole becomes extremal, with \(T=0\). In this extremal limit, the entropy and angular momentum, which are independent of $\lambda_{1,2}$, satisfy the same relation as in the Kerr black hole, \(S_{\mathrm{ext}}=2\pi J_{\mathrm{ext}}\). Note that the satisfaction of the first law does not require $\delta B=0$, implying that the parameter $B$ is intrinsically not a thermodynamic variable in our setup of black hole thermodynamics.

As discussed above in \eqref{remove CTC and conical 2}, the \(a\to0\) limit of \(ds_{\rm I,-}^2\) is highly nontrivial. Building on this, we identify the coordinate transformation \eqref{coord first law 1} that yields a consistent thermodynamic description. We now examine the static limit. We find that the mass and entropy vanish at order \(a\) and \(a^2\), respectively, while the angular momentum vanishes as \(a^5\). The temperature, on the other hand, diverges as \(a^{-1}\)
\bea
M&\sim&\frac{r_h \sqrt{\sqrt{1+B^2 r_h^2}-1}}{\sqrt{2} B \big(1+B^2 r_h^2\big)^{3/4}} a\,,\qquad
S\sim \frac{2 \pi  r_h^4}{(1+B^2 r_h^2)^{3/2} \big(1+\sqrt{1+B^2 r_h^2}\big)}a^2\,,\cr
 J&\sim& \frac{r_h }{4 \sqrt{1+B^2 r_h^2}}\Big[1+\frac{1}{\sqrt{1+B^2 r_h^2}}\Big]a^5\,,\qquad T\sim\frac{B (1+B^2 r_h^2)^{3/4}}{4 \sqrt{2} \pi  r_h \sqrt{\sqrt{1+B^2 r_h^2}-1}}\frac{1}{a}\,.
\eea
This singular limit is not surprising. As discussed, the global structure does not admit a smooth $a\rightarrow 0$ limit, since in the static solution $x=\pm 1$ are curvature singularities and are resolved only by rotations.

Finally, it is worth commenting that the absence of naked CTC provides restrictions on the three parameters $(B,\mu,a)$. However, the first law we obtained above does not appear to be affected by these restrictions. It still seems to be sensible to discuss the first law of black hole thermodynamics in the presence of naked CTCs. This is because the first law of black hole thermodynamics has a classical origin as the Iyer-Wald identity formulated in the covariant phase-space formalism \cite{Wald:1993nt,Iyer:1994ys}. It states that there must be a first-order differential relation among the integration constants evaluated on the horizon and at asymptotic infinity, provided that there is no singularity in the spacetime between the horizon and asymptotic infinity. (The situation can be more subtle \cite{Liu:2022wku} when there are Dirac or Misner strings \cite{Dirac:1931kp, Misner:1963fr}.) This is precisely why we can also establish the first law for the black hole inner horizon as well, even though the spacetime between the inner and outer horizons can involve a number of undesirable properties as CTCs. Thus, there must exist a first law even when naked CTCs exist, since there is no singularity in the spacetime for $r\in [r_h, \infty)$. It is thus the causality consideration rather than the first law that requires the absence of naked CTCs.

\subsection{Finite Ergoregions}

Having obtained the black holes with suitable $(t,\varphi)$ coordinates that yield thermodynamic quantities satisfying the first law and the absence of naked CTCs, it is still worth examining whether the ergospheres are finite or extended all the way to infinity, as in the case of Kerr-Melvin black hole, analysed in \cite{Gibbons:2013yq}. The black hole described by $ds_{\rm I,+}^2$ can have finite ergoregions \cite{Ma:2026otg}. For the metric $ds_{\rm I,-}^2$ after the transformations \eqref{remove CTC and conical} and \eqref{remove CTC and conical 2}, on the other hand, the region associated with the metric component $g_{tt}>0$ with a certain segment in $x\in [-1,1]$ extends all the way to asymptotic $r\rightarrow \infty$.

This in itself does not mean that the ergoregion is not finite. A thorough check requires us to consider a generic comoving coordinate $\phi\rightarrow \phi - \omega t$ and determine whether the resulting $g_{tt}>0$ region is restricted to some finite range of $r$. However, in order to derive the first law, we find that it is necessary to perform the coordinate transformation \eqref{coord first law 1}, which effectively fixes the comoving coordinate with $\omega = -\lambda_2 B$. It would then be highly nontrivial if the comoving coordinate fixed by the first law satisfies the requirement of a finite ergoregion. Intriguingly a miracle does occur in this case. We find that in the metric obtained after \eqref{coord first law 1}, the region of $g_{tt}>0$ is indeed confined to $r<r_*$ for some finite $r_*$ under suitable choices of the parameters $(B,\mu, a)$. Specifically, when $\mu\ge \mu_{\rm ext}>\mu_{\rm ctc}$, ergoregions generally extend to asymptotic infinity. However, when $\mu>\mu_{\rm ctc}> \mu_{\rm ext}$, ergoregions are generally finite. This particularly implies that the requirement of a finite ergoregion rules out extremal black holes described by the $ds_{\rm I, -}^2$ metric.

Before concluding, we present examples of black holes with finite ergoregions, given by
\be
B=1\,,\quad a=\frac25\,,\qquad \rightarrow\qquad \mu_{\rm max}^2=\frac{529}{525}> \mu_{\rm ctc}^2 = \frac{529}{105}(2\sqrt{21}-9)> \mu_{\rm ext}^2 = \frac{2116}{13125}\,.
\ee
For black holes free of naked CTCs, we must choose the parameter $\mu$ to lie between $\mu_{\rm max}$ and $\mu_{\rm ctc}$. We further find that there is a critical value $\tilde \mu\sim 0.913>\mu_{\rm ctc}\sim 0.912$ such that if $\mu>\tilde \mu$, the ergoregions do not extend to asymptotic infinity. As a specific example, we consider a black hole with $\mu=1>\mu_{\rm ctc}$, in which case the horizon radius $r_h=23(105+\sqrt{11009})/20\sim 241$. We find that the ergoregion is very close to the horizon, within $r<r_*$ with $r_*\sim 244$ for all $x\in [-1,1]$. In other words, we have $g_{tt}<0$ as long as $r>r_*$.
For this black hole, the thermodynamic quantities are
\be
M=0.301\,,\qquad J=0.00331\,,\qquad T=0.132\,,\qquad S=1.14\,,\qquad \omega_h=0.0304\,.
\ee
Note that these thermodynamic quantities match precisely those of a Kerr black hole, since we imposed the same CR relation when deriving the first law. Thus, our rotating metric
$ds_{\rm I,-}^2$ resolves the naked singularity of SLC metric and gives rise to a black hole with sensible thermodynamics, possesses no naked CTCs and ergoregions of finite extent.

\section{Solution II: local metric and branches}
\label{sec:Class II-local}

We now present a new Ricci-flat rotating metric, which we shall refer to as Solution II. This is our key finding, and in this section we focus on the discussion of its local metric, given by
\bea
ds_{\rm II}^2&=&\frac{(r^2-a^2)^2 (1-x^2) P-a^2 (1+x^2)^2 Q}{\mathbb{L}_{2}\Omega^4}\Big[d\varphi+\frac{a\mathbb{M}_{2}dt}{(r^2-a^2)^2 (1-x^2) P-a^2 (1+x^2)^2 Q}
\Big]^2\cr
&&+\mathbb{L}_{2}\Big[
\frac{1}{P}\frac{dx^2}{1-x^2}+\frac{dr^2}{Q}-\frac{(1-x^2)PQ}{(r^2-a^2)^2 (1-x^2) P-a^2 (1+x^2)^2 Q}dt^2
\Big]\,.\label{Ricci-flatII}
\eea
Although the metric is complicated, its determinant is simple, with $\sqrt{-g}=\mathbb{L}_{2}/\Omega^2$. Here, we have introduced two functions, \(\mathbb{L}_{2}\) and \(\mathbb{M}_{2}\), and a constant \(B_{0}\):
\cramp{
\bea
\mathbb{L}_2&=&\frac{4(1-B_0)^2}{B^4\Omega^2}\Big\{
\big[r(1-B_0+\Omega)-(r^2-a^2)\partial_r\Omega\big]^2+a^2
\big[x(1-B_0+\Omega)-(1+x^2)\partial_x\Omega\big]^2
\Big\}\,,\nn\\
\mathbb{M}_2&=&\frac{8(1-B_0)^3}{B^4}\Big[
P(r^2-a^2)(1-x^2)(1-B_0+\Omega-x\partial_x\Omega)+(1+x^2)
Q(1-B_0+\Omega-r\partial_r\Omega)\Big]\,,\nn\\
B_{0}&=&1-\sqrt{1-a^2 B^2 (1+B^2 \mu ^2)}\,.\label{M2L2B0}
\eea}
The remaining functions are given by
\bea
\Sigma&=&r^2+a^2x^2\,,\qquad P=1+B^2 x^2 \big[\mu ^2-a^2 (1+B^2 \mu ^2)^2\big]\,,\qquad Q=I\Delta\,,\cr
\Omega^2&=&I-B^2 x^2 \Delta\,,\qquad I=B^2 r^2+(1-B^2  \mu r)^2\,,\qquad \Delta= (r^2+a^2) (1+B^2 \mu ^2)-2  \mu r\,.
\label{GKBR solution}
\eea
As in Solution I, the function \(\Omega\) appearing in $ds_{\rm II}^2$ also has two branches when expressed in terms of $(r,x)$ coordinates, namely
\be
\Omega=\Omega_{\pm}=\pm\sqrt{I-B^2x^2\Delta}\,.\label{IIomegapm}
\ee
Furthermore, it can be verified that the constant parameter \(B_0\) in the metric also has two branches, i,e.,~\(B_{0,\pm}=1\pm\sqrt{1-a^2B^2(1+B^2\mu^2)}\). In other words, the new metric is Ricci flat, independent of the above four $\pm$ signs. Thus, a priori, Solution II appears to contain four different branches. However, we find that the solution sets \(\{\Omega_-,B_{0,-}\}\) and \(\{\Omega_+,B_{0,+}\}\) are identical, while \(\{\Omega_+,B_{0,-}\}\) and \(\{\Omega_-,B_{0,+}\}\) are also identical. Therefore, there are only two independent branches. Without loss of generality, we restrict our attention to the \(B_{0,-}\) branch, as presented in \eqref{M2L2B0}, so that $B_0$ vanishes when $B=0$. The two branches of the metric are denoted as
\be
ds_{\rm II,+}^2 = \fft{B^4}{16} ds_{\rm II}^2(\Omega_+)\,,\qquad
ds_{\rm II,-}^2 = ds_{\rm II}^2(\Omega_-)\,.
\ee
Note that a trombone rescaling was imposed so that both $ds_{\rm II,\pm}^2$ have a smooth $B\rightarrow 0$ limit. Analogous to Solution I, for nonvanishing $B$, the two branches are unified in the $(r,\Omega)$ coordinates; however, when $B=0$, we have $\Omega=1$, which no longer serves as a coordinate. In this case, the two branches become intrinsically distinct. We shall return to discuss the $B=0$ case in Section \ref{sec:B=0}.

In the static $a=0$ limit, neither of the metrics $ds_{\rm II,\pm}^2$ is new, but locally related to the static metrics $ds_{\rm I,\pm}^2$. Specifically, we find
\be
ds_{\rm II,\pm}^2(a=0,r_{\rm II} =\frac{r_{\rm I}}{1+B^2\mu r_{\rm I}})=ds_{\rm I,\pm}^2(a=0)\,,\label{staticrelation}
\ee
together with an additional transformation $\varphi_{\rm II,+} \rightarrow -16 \varphi_{\rm I,+}/B^4$ for the + branch. Furthermore, in the $B=0$ limit, both the $ds_{\rm I,+}^2$ and $ds_{\rm II,+}^2$ metrics reduce to the same Kerr metric. This raises the suspicion that $ds_{\rm I,+}^2$ and $ds_{\rm II,+}^2$ might be locally related. We now present strong evidence that they are not locally related.

Both metrics $ds_{\rm I,+}^2$ and $ds_{\rm II,+}^2$ are very complicated and distinguishing them using curvature invariants is not feasible. It is perhaps an even more daunting task trying to show that no such coordinate transformation exists. However, it is sufficient to show that such a coordinate transformation does not exist for small $B$, since the parameter $B$ is what makes $ds_{\rm I,+}^2$ and $ds_{\rm II,+}^2$ distinct from the same Kerr metric. We therefore seek coordinate transformations involving $(r,x)$ that map \(ds_{\rm II,+}^2\) to \(ds_{\rm I,+}^2\), working perturbatively in the parameter \(B\), namely
\be
r\quad\rightarrow\quad r'=r+B^2R(r,x) + {\cal O}(B^4)\,,\qquad x\quad\rightarrow\quad x'=x+B^2X(r,x) + {\cal O}(B^4)\,.
\ee
We require \(ds_{\rm I,+}^2\) and \(ds_{\rm II,+}^2\) to coincide under this coordinate transformation. As a necessary condition, we set the Killing coordinates $(t,\varphi)$ to be fixed and restrict the matching of the two metrics in the $(r,x)$ subspace only. This leads to three partial differential equations up to and including the quadratic order in $B$. They are $\{D_1,D_2,D_3\}=0$, with
\bea
D_1&=&(1-x^2) \big(a^4 x^2-4 \mu  r^3+a^2 r (r+2 \mu )\cr
&&-4 (r^2+a^2 x^2) X^{(0,1)}\big)-4 r (1-x^2) R - 4 (r^2+a^2) x X\,,\cr
D_2&=&a^6 x^2-4 r^4 (2 r-5 \mu ) \mu -4 a^2 x (r^2-2 r \mu +a^2) X+a^4 \big(r^2 (1+x^2)\cr
&&+2 r (1-4 x^2) \mu +2 x^2 \mu ^2\big)+a^2 r^2 \big(r^2-2 r (5+2 x^2) \mu -2 (1-6 x^2) \mu ^2\big)\cr
&&+4 \big(a^2 x^2 (r-\mu )+r (r \mu-a^2)\big) R-4 (r^2+a^2 x^2) (r^2-2 r \mu +a^2) R^{(1,0)}\,,\cr
D_3&=&(1-x^2) R^{(0,1)}+(r^2-2  \mu r +a^2) X^{(1,0)}\,.
\eea
Here $X^{(1,0)}$ and $X^{(0,1)}$ denote derivatives with respect to $r$ and $x$, respectively. It is clear that \(R\) can be solved for from \(D_1=0\) and completely eliminated from \(D_{2,3}\), reducing the system to two partial differential equations for \(X\). To examine whether this system admits a solution, we solve the remaining equations for \(X^{(0,2)}\) and \(X^{(1,1)}\). The consistency condition \(\partial_r X^{(0,2)}=\partial_x X^{(1,1)}\) then allows us to determine all second derivatives of \(X\), namely
\bea
X^{(0,2)}&=&\frac{1}{2 (1-x^2)^2 (r^2+a^2 x^2)}\Big[
a^4 x (1-x^2)^2-2 (r^2+a^2) (1+x^2) X-2 x (1-x^2) \big(r^2+a^2\cr
&&+2 a^2 (1-x^2)\big)
   X^{(0,1)}+2 r (1-x^2) (r^2-2 \mu  r+a^2) X^{(1,0)}
\Big]\,,\nn\\
X^{(1,1)}&=&-\frac{1}{4 r (r^2+a^2 x^2)}\Bigg\{
\frac{a^2 r^2 (r^2+2 r \mu +a^2)}{r^2-2 r \mu +a^2}+\frac{4 (r^2-a^2) x (r^2-3 r \mu +a^2)}{(1-x^2) (r^2-2 r \mu +a^2)}X\cr
&&-\frac{4 X^{(0,1)}-a^2}{r^2-2 r \mu +a^2}\Big[a^4 x^2-r^3 (r-3 \mu )-a^2 r \big(2 r (1-x^2)+3 x^2 \mu \big)\Big]\cr
&&\qquad\qquad+\frac{4 r (r^2+a^2) x }{1-x^2}X^{(1,0)}
\Bigg\}\,,\nn\\
X^{(2,0)}&=&-\frac{1}{2 (r^2+a^2 x^2) (r^2-2\mu r  +a^2)}\Big[
a^4 x (1-x^2)-2 a^2 (1+x^2) X-6 a^2 x (1-x^2) X^{(0,1)}\cr
&&+2 \big(r (a^2+r (2 r-3 \mu ))+a^2 x^2
   (r-\mu )\big) X^{(1,0)}
\Big]\,.\label{X2}
\eea
The consistency condition also requires \(\partial_r X^{(1,1)}=\partial_x X^{(2,0)}\). Using the above relations \eqref{X2}, we can then derive equations for the first derivative of \(X\):
\bea
X^{(0,1)}&=&-\frac{-a^2 (1-x^2) (5 r^4-24 a^2 r^2 x^2+3 a^4 x^4)+12 x \big(r^4+a^4 x^2-3 a^2 r^2 (1+x^2)\big) X}{12
   (1-x^2) (r^4-6 a^2 r^2 x^2+a^4 x^4)}\,.\label{X1}
\eea
The expression for \(X^{(2,0)}\) obtained above must be consistent with that given in \eqref{X2}. This consistency condition determines \(X^{(1,0)}\)
\bea
&&X^{(1,0)}=\frac{a^2 r}{3 (r^4-6 a^2 r^2 x^2+a^4 x^4)^2 (r^2-2 \mu  r+a^2)}\Bigg[
a^2 x (1-x^2) (9 r^6-37 a^2 r^4 x^2+47 a^4 r^2 x^4\cr
&&-3 a^6 x^6)+12 \Big[r^6 (1-2 x^2)+a^6 x^6
   (2-x^2)+a^2 r^4 x^2 (6+x^2)-a^4 r^2 x^4 (1+6 x^2)\Big] X
\Bigg].\label{X11}
\eea
Equations \eqref{X1} and \eqref{X11} yield an inconsistent expression for \(X^{(1,1)}\), so the system is not self-consistent and no solution exists.

An analogous argument can be made along a different route. When $B=0$, we find that $ds_{\rm I,+}^2$ and $ds_{\rm II,+}^2$ are locally related. It is the rotation parameter $a$ that makes the two metrics distinct. We can thus consider the coordinate transformation:
\be
r\rightarrow r' =\frac{r}{1+B^2\mu r} + a^2 R(r,x) + {\cal O}(a^4)\,,\qquad
x\rightarrow x'= x + a^2 X(r,x) + {\cal O}(a^4)\,.
\ee
In this case, the calculation is much more complicated since it involves a square root term involving $\Omega$. For simplicity, we consider the extremal case, in which $\mu$ is chosen so that $\Delta$ has a double root. We can then show that the equations are inconsistent, and hence no solution exists. We shall not present the details here, since the calculation is repetitive.

These do not constitute the most rigorous mathematical proof that $ds_{\rm I,+}^2$ and $ds_{\rm II,+}^2$ must be locally inequivalent when $B a \ne 0$, since coordinate transformations could in principle involve all the $(r,x,t,\varphi)$ coordinates, including the Killing coordinates. Nevertheless, the above demonstration, which restricts to the $(r,x)$ subspace while leaves out the Killing coordinates, provides strong evidence that the two metrics are locally inequivalent. In the next section, we shall analyse their global structures and show that distinctions not only persist, but also become more striking.

\section{Solution II: global structure and black hole thermodynamics}
\label{sec:Class II-global}

In this section, we analyse the global structures of both $ds_{\rm II,\pm}^2$ branches. Although they can be unified in the $(r,\Omega)$ coordinates as two different branches for nonzero $B$, their global structures are remarkably distinct. First, we note that for both branches, the reality condition for $B_0$ requires
\be
\mu \le \mu_{\rm real} \equiv \frac{\sqrt{1-a^2 B^2}}{a B^2}\,,\qquad \rightarrow\qquad aB<1\,.
\ee
This upper bound on $\mu$ diverges at the $a B\rightarrow 0$ limit, consistent with the known $a\rightarrow 0$ or $B\rightarrow 0$ limits.

\subsection{$ds_{\rm II,+}^2$: black holes with disjoint ergoregions}

This metric simply reduces to the Kerr metric when $B=0$. This is therefore another $B$-deformed Kerr metric. As discussed, the $a=0$ limit is locally equivalent to the $B$-deformed Schwarzschild black hole, up to a coordinate transformation given in \eqref{staticrelation}. The metric has a total of four degenerate surfaces. Two of them are given by $x=\pm 1$ and the other two are associated with the two roots of the quadratic function $\Delta(r)$, i.e.:
\be
r_\pm = \frac{\mu \pm \sqrt{\mu^2 - a^2 (1+B^2 \mu^2)^2}}{1 + B^2 \mu^2}\,.
\ee
The reality condition requires that we have
\be
\mu_{\rm ext}^- \le \mu \le \mu_{\rm ext}^+\,,\qquad \mu_{\rm ext}^\pm = \fft{1\pm \sqrt{1-4a^2 B^2}}{2aB^2}\,.
\ee
An intriguing new feature that does not exist in other known black holes is that $\Delta(r)$ has double roots for both $\mu^\pm_{\rm ext}$. Furthermore, it can be readily verified that the two values $\mu^\pm_{\rm ext}$ cannot be equal. We can thus have two extremal solutions for a given choice of $(a,B)$ satisfying
\be
aB< \frac12\,,
\ee
where saturation is disallowed. It is easy to check that for the above parameter constraint, we always have $\mu_{\rm ext}^+<\mu_{\rm real}$.

We now consider the coordinate singularities at $x=\pm1$, which imply that a spacelike Killing vector becomes degenerate. For \(ds_{\rm II,+}^2\) in \eqref{Ricci-flatII}, since \(x=\pm1\) are not zeros of \(\mathbb{M}_{2,+}\), we can remove the manifest CTCs and conical singularities using the coordinate transformation \eqref{remove CTC and conical}. In this case,  \(\{\alpha_+,P_{0,+}\}\) are given by:
\be
\alpha_+=\frac{aB^4}{8(1-B_0)^4}\,,\qquad P_{0,+}=-\frac{B^4(1+B^2\mu^2)}{16(1-B_0)^2}\,.
\ee
Both quantities have smooth $a=0$ and/or $B=0$ limits. To investigate the possibility of naked CTCs, we examine the metric component associated with the periodic $\varphi$ coordinate, namely
\bea
g_{\varphi\varphi}&=&\frac{1}{4 (1-B_0)^4 (1+B^2 \mu ^2)^2 \big[(r^2-a^2)^2 (1-x^2) P-a^2 (1+x^2)^2
   Q\big]}\Big[
   -a^2 B^4 (1-x^2) P Q \mathbb{L}\cr
   &&+\frac{\big[a^2 B^4 \mathbb{M}+8 (1-B_0)^4 \big((r^2-a^2)^2 (1-x^2) P-a^2 (1+x^2)^2
   Q\big)\big]^2}{B^4  \Omega ^4\mathbb{L}}
   \Big].
\eea
We find that for a given $r\ge r_h$, $g_{\varphi\varphi}(x\in[-1,1])\ge 0$, where saturation occurs at $x=\pm 1$, associated with the two degenerate surfaces that require $\varphi$ to have a period of $2\pi$ to avoid a conic singularity. The maximum occurs at $x=0$ for all $r$; here we give its value on the horizon
\be
g_{\varphi\varphi}(r_h,x=0)=\frac{4\mu^2(1-B_0)^2}{(1+B^2\mu^2)^2}\,.
\ee
As $r$ increases, $g_{\varphi\varphi}(r,x=0)$ increases monotonically with $r$. For any given $r>r_h$, $g_{\varphi\varphi}$ is always positive except when it vanishes at $x=\pm1$; therefore, the solution has no naked CTCs.

We are now in a position to derive the thermodynamic properties associated with the coordinates \eqref{coord first law 1}, and we find
\bea
T&=&\lambda _1\frac{ (1+B^2 \mu ^2)^2 \sqrt{\mu ^2-a^2 (1+B^2 \mu ^2)^2} (1-B_0)^2}{4 \pi  \mu
   \big[\mu +\sqrt{\mu ^2-a^2 (1+B^2 \mu ^2)^2}\big]}\,,\cr
\omega_H&=&\lambda _1\frac{ a (1+B^2 \mu ^2)^3 (1-B_0)^2}{2 \mu  \big[\mu +\sqrt{\mu ^2-a^2 (1+B^2 \mu
   ^2)^2}\big]}-B \lambda _2\,,\qquad J=\frac{ \mu a}{(1+B^2 \mu ^2)^2}\,,\cr
   M&=&\lambda _1\frac{ \mu  (1-B_0)^2}{1+B^2 \mu ^2}-2\lambda_2BJ\,,\qquad
   S=\frac{2 \pi  \mu  \big[\mu +\sqrt{\mu ^2-a^2 (1+B^2 \mu ^2)^2}\big]}{(1+B^2 \mu ^2)^3}\,.
\eea
Similarly, imposing the first law and the CR relation determines \(\lambda_{1,2}\) as:
\be
\lambda_1=\frac{1}{(1-B_0)^2\sqrt{1+B^2\mu^2}}\,,\qquad \lambda_2=0\,.
\ee
It can be readily verified that the first law and CR relation in \eqref{firstlawCR} are satisfied. In the extremal limit, the entropy and angular momentum of the black hole, which are independent of $\lambda_{1,2}$, also satisfy the same relation as in the Kerr black hole, \(S_{\mathrm{ext}}=2\pi J_{\mathrm{ext}}\).

We now present two explicit examples. In the first example, we consider $(B,a)=(1,1/10)$. In this case, the allowed $\mu$ must lie in $[\mu_{\rm ext}^-, \mu_{\rm ext}^+]=[0.1010,9.899]$. Specifically, we set $\mu=1$, for which $r_h=0.9899$. The thermodynamic quantities of this black hole are:
\be
M=0.3536\,,\qquad J=0.025\,,\qquad S=1.555\,,\qquad T=0.1114\,,\qquad \omega=0.1429\,.
\ee
In the second example, we consider $B=1/10$ and $a=1$. The parameter $\mu$ then must lie within $[1.010, 98.99]$ for the solution to describe a black hole. For a specific choice of $\mu=5$, we have $r_h=7.873$, and the thermodynamic quantities are:
\be
M=3.578\,,\qquad J=3.200\,,\qquad S=158.3\,,\qquad T=0.01094\,,\qquad \omega=0.01775\,.
\ee
However, all the black holes that satisfy the first law have non-compact ergoregions. To see this, we note that:
\bea
g_{tt}(r\rightarrow\infty,x)&=&\frac{1}{4(1-x^2)}+
\frac{a^4B^4(1+B^2\mu^2)^2+2a^2B^2-1}{8a^2B^2(1+B^2\mu^2)(1-B_0)^2} +
{\cal O}(1-x^2)\,,\nn\\
g_{tt}(r,x^2=1)&=& -\frac{(1-B_0)^2}{a^2B^2 (1 + B^2 \mu^2)} + {\cal O}(r^{-2})\,.
\eea
On the one hand, $g_{tt}$ can be negative when we fix $x^2=1$ in the large $r$ limit. On the other hand, for $x^2\ne 1$, the quantity $g_{tt}$ diverges to positive infinity at $x^2\rightarrow 1$ at asymptotic infinity, indicating that ergoregions extend all the way to asymptotic infinity in the vicinity of the north and south poles. Such unusual behavior indicates that there exist disjoint ergoregions.

To illustrate the disjoint ergoregions, we use the second example of the black hole discussed above. We determine both upper and lower boundaries of the radial coordinate
\be
\{B,a,\mu\}=\{\frac{1}{10},1,5\}:\qquad r_1=8.125\,,\qquad r_2=200.01\,,
\ee
such that $g_{tt}$ is always negative when $r\in (r_1,r_2)$ for all $x\in [-1,1]$. As expected, ergoregions develop for $r\in (r_h,r_1)$ when $x\in (-\delta_0,\delta_0)$ with some $r$-dependent $\delta_0<1$. For example, when $r=8$, we have $\delta_0=0.5898$. New disjoint ergoregions develop when $r>r_2$, within two narrow bands close to the north and south poles, i.e. $x\in (-\delta_2,-\delta_1)$ and $x\in (\delta_1,\delta_2)$. By ``narrow'' bands, we mean that both $\delta_1$ and $\delta_2$ are very close to 1, associated with the north and south poles. For example, we consider $r=250>r_2$, in which case, two ergoregions lie within bands specified by $\delta_1=0.9868$ and $\delta_2=0.9995$. As $r$ increases, the lower bound $\delta_1$ approaches asymptotically to $\delta_1=0.9595$ while the upper bound $\delta_2$ approaches 1 but never reaches it.

Thus, we see that for suitable parameters, there exist radial regions where $g_{tt}<0$ for all points of the foliating spheres. But these black holes have two disjoint ergoregions. The near ergoregion encloses the equator, while the far ergoregions form narrow bands near the north and south poles but not including them. Of course, this analysis does not examine whether there exists another comoving coordinate so that only the ergoregion of finite extend exists. However, if such a comoving coordinate does exist, it is not compatible with the coordinate system based on which the first law was derived.

\subsection{$ds_{\rm II,-}^2$: black hole time machine}

Here we analyse the \(ds_{\rm II,-}^2\) branch, which has very unusual global properties. At first sight, the metric is similar to the previous examples,  having four degenerate surfaces, associated with $x=\pm 1$ and the two roots $r_\pm $ of $\Delta(r)$, which we analysed in the previous subsection. What is unusual is that it is the same Killing vector $\ell=\partial_t$ that becomes degenerate at all of $x=\pm 1$ and $r_\pm$. The surface gravity at $r_\pm $ is real, indicating that these are the inner and outer horizons, with temperatures
\be
T_{\pm}=\frac{(1+B^2 \mu ^2)^2 (1-B_0)^2}{4 \pi  \big[\mu \pm\sqrt{\mu ^2-a^2 (1+B^2 \mu ^2)^2}\big]}\,.\label{tem at horizon}
\ee
On the other hand, the surface gravity at $x=\pm 1$ is purely imaginary,  requiring the {\it real} time coordinate to be periodic with the (same) period:
\be
\Delta t= \frac{4 \pi  a}{(1+B^2 \mu ^2) (1-B_0)^2}\,.
\ee
The coordinate $\varphi$, on the other hand, is no longer required to be periodic. The fact that the real $t$ is required to be periodic to avoid a conical singularity at a degenerate surface was also observed in some over-rotating black holes \cite{Gibbons:1999uv,Cvetic:2005zi, Feng:2016dbw,Cvetic:2018ipt}, which are referred to as time machines.  The unusual aspect of our time machine is that there also exists an event horizon, which usually does not exist in any previously known time machine. We therefore refer to the $ds_{\rm II,-}^2$ metric as a black hole time machine.

The static limit of \(ds_{\rm II,-}^2\) coincides with the \(ds_-^2\) solution in \eqref{two static Ricci flat}. In other words, the static limit of \(ds_{\rm II,-}^2\) corresponds to a \(B\)-deformation of the SLC solution, implying that \(x=\pm1\) are curvature singularities in the \(a=0\) limit. In the rotating case, \(x=\pm1\) are regular surfaces. The expression for the squared Riemann tensor at $x=\pm 1$ is given by
\bea
\mathrm{Riem}^2&\sim&\frac{3(1-B_0)^8}{64a^6\Delta^6}\Big[
16 a^2 \mu ^4-48 a^2 r \mu ^3 (1+B^2 \mu ^2)-12 a^2 (a^2-5 r^2) \mu ^2 (1+B^2 \mu ^2)^2\cr
&&+8 a^2 r (3 a^2-5
   r^2) \mu  (1+B^2 \mu ^2)^3+(a^6-15 a^4 r^2+15 a^2 r^4-r^6) (1+B^2 \mu ^2)^4\,,
   \Big]\label{Riem2 2m x1}
\eea
which diverges on the horizons \(r=r_\pm\). For fixed $r=r_\pm$, the squared Riemann tensor is
\bea
\mathrm{Riem}^2&\sim&\frac{3 \big[(1+3 x^2)^2 r_\pm^2-a^2 x^2 (3+x^2)^2\big] \big[\mu  (r_\pm^2-a^2)^2+2 a^2 r_\pm
   (r_\pm^2+a^2)\big]^4}{(1-x^2)^6 \mu ^6 (r_\pm^2-a^2)^8 (r_\pm^2+a^2)^6}\,,
\eea
which diverges at $x=\pm 1$. Thus, the spacetime of the black hole time machine is singular when two surfaces degenerate simultaneously, namely $(r,x)=(r_\pm,1)$ or $(r,x)=(r_\pm,-1)$.

\section{ The $B=0$ limit}
\label{sec:B=0}

As was discussed, both Solutions I and II can have $\pm$ branches that can be locally unified in the $(r,\Omega)$ coordinate system for nonzero $B$. When $B=0$, the variable $\Omega$ reduces to 1 and hence ceases to be a coordinate. The ``$+$'' and ``$-$'' branches become locally distinct. When $B=0$, the metrics $ds_{\rm I,+}^2$ and $ds_{\rm II,+}^2$ become the same Kerr black hole with parameters $(\mu, a)$, associated with the mass $M=\mu$ and angular momentum $J=\mu a$. On the other hand, we shall show that the metrics $ds_{\rm I,-}^2$ and $ds_{\rm II,-}^2$ are distinct when $B=0$, unless we further set $a=0$, in which case, both metrics reduce to the same SLC metric. The two metrics with nonzero parameter $a$ can be viewed as rotating SLC metrics, which we refer to as R-SLC-I and R-SLC-II metrics, respectively. They are given by
\be
ds_{\rm R-SLC-I}^2 = ds_{\rm I,-}^2\Big|_{B=0}\,,\qquad
ds_{\rm R-SLC-II}^2 = ds_{\rm II,-}^2\Big|_{B=0}\,.
\ee
Note that both metrics $ds_{\rm I,-}^2$ and $ds_{\rm II,-}^2$ have a smooth $B\rightarrow 0$ limit. Specifically, the two R-SLC metrics are
\bea
&&ds_{\rm R-SLC-I}^2 =\tilde{\mathbb{L}}_1\Big(\frac{dr^2}{Q}+\frac{dx^2}{1-x^2}\Big)- \frac{(1-x^2)Q\tilde{\mathbb{L}}_1}{r^4(1-x^2)-a^2x^4Q}dt^2\cr
&&+\frac{r^4(1-x^2)-a^2x^4Q}{\tilde{\mathbb{L}}_1}\Big(d\varphi- \frac{a\tilde{\mathbb{M}}_1}{r^4(1-x^2)-a^2x^4Q}dt\Big)^2\,,\nn\\
&&ds_{\rm R-SLC-II}^2 =\tilde{\mathbb{L}}_2\Big(\frac{dr^2}{Q}+\frac{dx^2}{1-x^2}\Big)- \frac{(1-x^2)Q\tilde{\mathbb{L}}_2}{(r^2-a^2)^2(1-x^2)-a^2(1+x^2)^2Q}dt^2\cr
&&+\frac{(r^2-a^2)^2(1-x^2)-a^2(1+x^2)^2Q}{\tilde{\mathbb{L}}_2}\Big(d\varphi+ \frac{a\tilde{\mathbb{M}}_2}{(r^2-a^2)^2(1-x^2)-a^2(1+x^2)^2Q}dt\Big)^2\,,\nn
\eea
where
\bea
&&\tilde{\mathbb{L}}_1=r^6 (1-x^2)^2+a^2 x^2 (x^4 Q^2-4\mu r^3 x^2  +3 r^2 \Sigma )\,,\cr
&&\tilde{\mathbb{M}}_1=r^6 (1-x^2)+x^2 Q \Big[a^4 x^2+(1-x^2) \big(3 r^4-x^2 (a^4-3 a^2 r^2+2 \mu  r^3)\big)\Big]\,,\nn\\
&&\tilde{\mathbb{L}}_2=16 a^2 Q^2-\frac{2 a^2 Q (r^2-a^2)^2 (1-x^2)^2}{r^2}+\frac{(r^2-a^2)^4 (1-x^2)^2}{r^2}\cr
&&\qquad\qquad -\frac{a^2 Q^2 (1-x^2) }{r^2}\big[r^2
   (16+7 x^2+x^4)-a^2 (1-x^2)\big]\,,\cr
&&\tilde{\mathbb{M}}_2=\frac{Q (1-x^2)}{r^2}\Big[
2 Q r^2 (r^2-a^2) (1+3 x^2)+(r^2-a^2)^3 (1-x^4)\cr
&&\qquad\qquad +Q (1-x^4) \big(a^4-r^4-2 a^2 r (r-\mu )\big)
\Big]\,.
\eea
The functions $Q$ and $\Sigma$ are identical to those in the Kerr black hole, given by
\be
Q=r^2 - 2\mu r + a^2\,,\qquad \Sigma = r^2 + a^2 x^2\,.\label{QSigma}
\ee
It turns out that a rotating SLC metric was already constructed in \cite{Barrientos:2025rjn}, and is referred to as the Kerr-Levi-Civita (KLC) spacetime. The metric is
\bea
ds_{\mathrm{KLC}}^2&=&f(d\varphi-\omega dt)^2-\frac{Q(1-x^2)dt^2}{f}+\frac{(1-x^2)\big[(r^2+a^2)^2-a^2Q(1-x^2)\big]}{f}\Big(\frac{dr^2}{Q}+\frac{dx^2}{1-x^2}\Big)\,,\cr
f&=&\frac{(1-x^2)\Sigma\big((r^2+a^2)^2-a^2Q(1-x^2)\big)}{4\mu^2a^2x^2\big[a^2(1-x^2)^2+\Sigma(3-x^2)\big]^2+(1-x^2)^2\big[(r^2+a^2)^2-a^2Q(1-x^2)\big]^2}\,,\cr
\omega&=&-\frac{2\mu a}{a^2Qx^2+r(r^3+a^2r+2\mu a^2)}\Big[
(r^3-3a^2r+2\mu a^2)Qx^4-6r(r^2+a^2)Qx^2\cr
&&+(r^3+a^2r+2\mu a^2)(a^2-6\mu r-3r^2)
\Big]\,.\label{Kerr LC0}
\eea
where $(Q,\Sigma)$ are also quantities given in \eqref{QSigma}.

When $a=0$, all three metrics $ds_{\rm R-SLC-I}^2$, $ds_{\rm R-SLC-II}^2$ and  $ds_{\rm KLC}^2$ reduce to the same static SLC metric. Therefore, there are now three rotating generalizations of the SLC metric. It is natural to ask whether these three metrics are distinct or locally equivalent. Their distinctive global structures suggest that they are not locally equivalent. For example, the metric $ds_{\rm R-SLC-II}^2$ continues to describe a black hole time machine, as in the case of the original $B\ne 0$ solution. This argument is suggestive, but not sufficient to establish that they are locally distinct. However, the metrics are too complicated to compare their curvature tensors directly.

We present a stronger argument that these metrics are distinct in that they are not related by a coordinate transformation restricted to the $(r,x)$ coordinates, without involving the Killing coordinates. We begin by comparing the $ds_{\rm R-SLC-I}^2$ and $ds_{\rm KLC}^2$ metrics. We perform a transformation of the coordinates \(\{r,x\}\) for \(ds_{\rm R-SLC-I}^2\):
\be
r\quad \rightarrow \quad r'=r+a^2R(r,x)+\mathcal{O}(a^4)\,,\qquad x\quad \rightarrow\quad x'=x+a^2X(r,x)+\mathcal{O}(a^4)\,.
\ee
After expanding in \(a\) and letting the resulting two-dimensional metric be the same as that of $ds_{\rm KLC}^2$, we find that the two functions $(R,X)$ must satisfy three partial differential equations $D_1=0$, $D_2=0$ and $D_3=0$, with
\bea
D_1&=&3 r^3 R (1-x^2)^2-r^4 x (1-x^2) X+r^4 (1-x^2)^2 X^{(0,1)}-r^2 (1-3 x^2)\cr
&&-2 r (1-3 x^2+4 x^4) \mu -6 x^2 (3-2 x^2)
   \mu ^2\,,\cr
D_2&=&r^3 (1-x^2)^2 (2 r-5 \mu ) R+(r-2 \mu ) \big[r^4 (1-x^2)^2 R^{(1,0)}-2 r^4 x (1-x^2) X\cr
&&-r^2 (1-3 x^2)-2 r (1-3 x^2+4 x^4)
   \mu -6 x^2 (3-2 x^2) \mu ^2\big]\,,\cr
D_3&=&(1-x^2) R^{(0,1)}+r (r-2 \mu ) X^{(1,0)}\,.
\eea
We can eliminate \(R\) from the \(D_{2,3}\) equations by using the expression for \(R\) obtained from \(D_1\). Imposing the consistency conditions, we then obtain expressions for the second derivatives of \(X\) analogous to \eqref{X2}. We have
\bea
X^{(0,2)}&=&-\frac{1}{r^4 (1-x^2)^3}\Big[
2 x \big[r^2 (1+3 x^2)+2 r (1-5 x^2) \mu -6 (3-x^2) \mu ^2\big]\cr
&&-r^4 (1-x^4) X-r^4 x (1-x^2)^2 X^{(0,1)}-3 r^4
   (1-x^2)^2 (r-2 \mu ) X^{(1,0)}
\Big]\,,\cr
X^{(1,1)}&=&\frac{1}{r^5 (1-x^2)^2 (r-2 \mu )}\Big[
r^5 x (1-x^2) (r-2 \mu ) X^{(1,0)}-r^4 (1-x^2)^2 (3 r-7 \mu ) X^{(0,1)}\cr
&&-r^4 x (1-x^2)  (3 r-5 \mu )X-2 r^3 (1-3 x^2)-3 r^2 (1-3
   x^2+8 x^4) \mu\cr
&& +2 r (5-51 x^2+44 x^4) \mu ^2 +42 x^2 (3-2 x^2) \mu ^3
\Big]\,,\cr
X^{(2,0)}&=&\frac{1}{r^5 (1-x^2)^2 (r-2 \mu )}\Big[
2 x \big(r^2 (1+3 x^2)+2 r (1-5 x^2) \mu -6 (3-x^2) \mu ^2\big)\cr
&&-2 r^4 (1-x^4) X -2 r^4 x (1-x^2)^2 X^{(0,1)}-r^4
   (1-x^2)^2 (4 r-7 \mu ) X^{(1,0)}
\Big]\,.
\eea
Likewise, by imposing the consistency conditions, we can reduce the above three equations to two equations involving the first derivatives of \(X\):
\bea
X^{(0,1)}&=&-\frac{1}{r^4 (1-x^2)^2 \big(4 r-7 (1-x^2) \mu \big)}\Big[
2 r^3 (1-21 x^2)+r^2 (5+8 x^2+139 x^4-72 x^6) \mu\cr
&&-4 r (1-139 x^2+196 x^4-88 x^6) \mu ^2-210 x^2 (3-5 x^2+2 x^4) \mu ^3\cr
&&+r^4 x (1-x^2)  \big(8 r-5 (1-x^2) \mu \big)X\Big]\,,\nn\\
X^{(1,0)} &=& \frac{1}{r^4 (1-x^2)^2 (r-2 \mu ) \big(4 r-7 (1-x^2) \mu \big)^2}\Big[
24 r^4 x (5-x^2)-4 r^3 x (55+240 x^2-367 x^4\cr
&&+120 x^6) \mu -8 r^2 x (203-840 x^2+1093 x^4-456 x^6+36 x^8) \mu ^2\cr
&&+4 r x (1-x^2)
   (1243-2432 x^2 +1883 x^4-352 x^6) \mu ^3 +(7-5 x^2-2 x^4) \mu ^2\big) X\cr
&&-168 x (1-x^2)^2 (21-22 x^2+10 x^4) \mu ^4-4 r^4 (1-x^2)^2 \big(4 r^2-r (11+4
   x^2) \mu\Big]\,.
\eea
Furthermore, imposing the consistency conditions allows us to obtain an expression for \(X\). However, we find that the resulting expression is not self-consistent, indicating that no solution exists. We perform the same analysis for the pairs ($ds_{\rm R-SLC-II}^2$, $ds_{\rm KLC}^2$) and ($ds_{\rm R-SLC-I}^2$, $ds_{\rm R-SLC-II}^2$) and draw the same conclusion. These analyses are not yet conclusive in showing that these three metrics are locally inequivalent; however, if they are locally equivalent, the coordinate transformation must involve all the four $(r,x,t,\varphi)$ coordinates. Together with the fact that these metrics have totally distinct global properties, this suggests that they should be treated as different metrics.

\section{Magnetizing the Ricci-flat seeds}
\label{sec:magnetization}

In this section, we consider remagnetizing the Ricci-flat seeds discussed earlier. We consider two theories: one is Einstein-Maxwell gravity and the other is KK theory. In both cases, the magnetization can be achieved by a solution-generating technique from a seed.

\subsection{Einstein-Maxwell theory}

Einstein-Maxwell gravity reduced on a circle gives rise to $D=3$ gravity coupled to a scalar coset $SU(2,1)/(SU(2)\times U(1))$. The magnetization of a four-dimensional Ricci-flat seed involves a specific $SU(2,1)$ global transformation, given by \eqref{Magnetizing U}. Since both of our metrics $ds_{\rm I}^2$ and $ds_{\rm II}^2$ already have three parameters $(B,\mu,a)$, the resulting general metrics will have four parameters $({\cal B}, B, \mu, a)$. The general solutions are very complicated; however, setting $B=0$ yields the known Kerr-Melvin black hole for both $ds_{\rm I,+}^2$ and $ds_{\rm II,+}^2$, where ${\cal B}$ is the magnetic-field parameter.

As was shown in \cite{Ma:2026otg}, for $ds_{\rm I}^2$, another miraculous simplification occurs when ${\cal B}=B$. The key simplification is that the function $\Omega$ appears in the new metric only via $\Omega^2$. The linear term in $\Omega$ appears in the Maxwell field only in such a way that the sign change $\Omega\rightarrow -\Omega$ implies $A\rightarrow -A$. In other words, both $ds_{\rm I,\pm}^2$ branches yield the same solution, the KBR solution \cite{Podolsky:2025tle,Ovcharenko:2025cpm}. Note that this statement is only true for ${\cal B}=B\ne 0$. The seed metrics $ds_{\rm I,\pm}^2$ become distinct if we take the $B\rightarrow 0$ limit before magnetization and hence the resulting magnetizations also yield distinct results.

An analogous situation occurs for the $ds_{\rm II}^2$. In this case, the miraculous simplification occurs when we take
\be
\mathcal{B}=\frac{4(1-B_0)}{B}\,.
\ee
The resulting magnetized rotating black hole solution in Einstein-Maxwell gravity is
\bea
ds^2&=&\frac{16(1-B_0)^4}{B^4\Omega^2}\Bigg\{
-\frac{Q}{\Sigma}\Big[dt-\frac{aB^4}{16(1-B_0)^4}(1+x^2)d\varphi\Big]^2
+\frac{\Sigma}{Q}dr^2+\frac{\Sigma}{P}\frac{dx^2}{1-x^2}\cr
&&+\frac{P}{\Sigma}(1-x^2)\Big[adt+\frac{B^4}{16(1-B_0)^4}(r^2-a^2)d\varphi\Big]^2
\Bigg\}\,,\cr
A_{\1}&=&\frac{4(1-B_0)^2}{B^3}\Bigg\{
\frac{r\partial_r\Omega}{\Sigma}\Big[adt+\frac{B^4}{16(1-B_0)^4}(r^2-a^2)d\varphi\Big]\cr
&&-\frac{ax\partial_x\Omega}{\Sigma}\Big[dt-\frac{aB^4}{16(1-B_0)^4}(1+x^2)d\varphi\Big]
-\frac{B^4}{16(1-B_0)^4}(\Omega-1)d\varphi
\Bigg\}\,.\label{GKBR}
\eea
This solution is not yet quite the same as the genuine KBR black hole, since it appears to diverge in both the \(a\to0\) and \(B\to0\) limits. In order to obtain the exact G-KBR solution, we find that a trombone symmetry transformation is necessary, accompanied by an appropriate rescaling of the azimuthal coordinate \(\varphi\):
\be
g_{\mu\nu}\rightarrow g_{\mu\nu}'=\frac{B^4}{16(1-B_0)^4}g_{\mu\nu}\,,\quad A_\mu\rightarrow A_{\mu}'=\frac{B^2}{4(1-B_0)^2}A_{\mu}\,,\quad \varphi\rightarrow\varphi'=-\frac{16(1-B_0)^4}{B^4}\varphi\,,
\ee
which then brings the solution into the form of the G-KBR solutions obtained in \cite{Ovcharenko:2026tos,Kubiznak:2026uro}. In this form, the solution possesses well-defined static and zero-magnetic-field limits. By performing a further transformation of the time coordinate, namely $t\rightarrow t'=t - 2 a \varphi$, the Killing vector $\partial_\varphi$ becomes degenerate at $x=\pm 1$.

We now consider a new magnetization. It is readily seen that both metrics $ds_{\rm I}^2$ and $ds_{\rm II}^2$ remain real under the replacement \(B=\mathrm{i}E\). Therefore, we can magnetize the Ricci-flat metrics with paramters $(E,\mu,a)$. This was done for
$ds_{\rm I,+}^2$ in \cite{Ma:2026ima} and we now follow the same procedure to magnetize $ds_{\rm II,+}^2$, with $B_0$ now replaced by
\be
E_0=1-\sqrt{1+a^2 E^2 (1-E^2 \mu ^2)}\,.
\ee
The resulting solution admits a smooth static limit \(a\to0\). However, it becomes singular in the zero-field limit \(E\to0\). To obtain a well-defined limit, we perform an appropriate trombone rescaling, accompanied by a reparameterization of the azimuthal coordinate \(\varphi\),
\be
g_{\mu\nu}\rightarrow g_{\mu\nu}'=\frac{E^4}{16}g_{\mu\nu}\,,\qquad A_\mu\rightarrow A_{\mu}'=\frac{E^2}{4}A_{\mu}\,,\qquad \varphi\rightarrow\varphi'=\frac{16}{E^4}\varphi\,,
\ee
thereby arriving at an exact rotating black hole solution immersed in an external electric or magnetic field. The metric is
\bea
ds^2&=&-\frac{Q}{\mathbf{L} \Omega ^4}\Big[
a (1+x^2) d\varphi+\frac{a^2 (1+x^2) \mathbf{H}+(r^2-a^2) (1-x^2) P \mathbf{L}
   \Omega ^2 }{(r^2-a^2)^2 (1-x^2) P-a^2  (1+x^2)^2Q}dt
\Big]^2 \cr
&&+\frac{P (1-x^2)}{\mathbf{L} \Omega ^4}\Big[
(r^2-a^2) d\varphi+\frac{a \big((r^2-a^2) \mathbf{H}+(1+x^2) Q \mathbf{L} \Omega
   ^2\big) }{(r^2-a^2)^2 (1-x^2) P-a^2 (1+x^2)^2 Q}dt
\Big]^2\cr
&&+\mathbf{L} \Big(\frac{dr^2}{Q}+\frac{1}{P}\frac{dx^2}{1-x^2}\Big)\,.\label{E-M field}
\eea
with the determinant $\sqrt{-g}=\mathbf{L}/\Omega^2$. Here we have introduced two new functions $\{\mathbf{H},\mathbf{L}\}$
\bea
\mathbf{L}&=&\frac{(1-E_0)^2}{\Omega ^2}\Big[
\big(r\Omega-(r^2-a^2)\partial_r\Omega\big)^2+a^2\big(x\Omega-(1+x^2)\partial_x\Omega\big)^2
\Big]\,,\cr
\mathbf{H}&=&(1-E_0)^4 \Big[(r^2-a^2)(1-x^2)P+(1+x^2)Q
 \Big]\,.\label{LH}
\eea
All the remaining functions are obtained from those in \eqref{GKBR solution} by setting \(B=\mathrm{i} E\). They are
\bea
\Sigma&=&r^2+a^2x^2\,,\qquad P=1-E^2 x^2 \big[\mu ^2-a^2 (1-E^2 \mu ^2)^2\big]\,,\qquad Q=I\Delta\,,\cr
\Omega^2&=&I+E^2 x^2 \Delta\,,\quad I=-E^2 r^2+(1+E^2  \mu r)^2\,,\quad \Delta= (r^2+a^2) (1-E^2 \mu ^2)-2  \mu r\,.
\eea
The Maxwell field is given by
\bea
A_{\1}^{\mathrm{ele}}&=&-\frac{a E (1-E_0)^2 x}{\mathbf{L} \Omega ^3}\Big[
d\varphi+\frac{a\mathbf{H}dt}{(r^2-a^2)^2 (1-x^2) P-a^2  (1+x^2)^2Q}
\Big]\Big[2 r (1-E^2 \mu ^2) \Sigma\cr
&& +\mu a^2 (1-x^2)-\mu r^2 (3+x^2) \Big]-\frac{dt  (1-E_0)^2(1-x^2)\partial_x\Omega}{E\big[(r^2-a^2)^2 (1-x^2) P-a^2  (1+x^2)^2Q\big]}\Big[
\cr
&& a^2 (1+x^2) \mu  -3 a^2 E^2 r^2 (1+x^2) \mu  (1-E^2 \mu ^2)-r^3 (1-E^2 x^2 \mu
   ^2)\cr
&&+a^2 r \big(3 E^2 \mu ^2-x^2 (1-2 E^2 \mu ^2)\big) +a^2 E^2 r (1-E^2 \mu
   ^2)^2 \Sigma
\Big]\,,\cr
A_{\1}^{\mathrm{mag}}&=&\frac{(1-E_0)^2}{E \mathbf{L} \Omega ^3}\Big[
d\varphi+\frac{a\mathbf{H}dt}{(r^2-a^2)^2 (1-x^2) P-a^2  (1+x^2)^2Q}
\Big]\Big[\cr
&& E^2 r \big(r^2 (1-x^2)+a^2 (1+3 x^2)\big) \mu +\big(1-a^2 E^2 (1-E^2 \mu ^2)\big)
   \Sigma\Big]\cr
&&+\frac{a dt}{E (1-E_0)^2 \Omega  \big[ (r^2-a^2)^2 (1-x^2)P-a^2
   (1+x^2)^2Q\big]}\Big[I\mathbf{H}\cr
   &&-E^2(1-E_0)^4(1+x^2)Q\big[\mu r(1+x^2)-(1-E^2\mu^2)\Sigma\big]\Big].
\eea
To eliminate the naked CTCs and conical singularities, we perform the coordinate transformation
\bea
\varphi\quad \rightarrow\quad \varphi'=\frac{\varphi}{P_0}\,,\qquad t\quad \rightarrow \quad t'=t+\frac{2a}{(1-E_0)^4}\frac{\varphi}{P_0}\,,\qquad P_0=\frac{1-E^2\mu^2}{(1-E_0)^2}\,.
\eea
We can readily verify that this solution carries neither electric nor magnetic charges, i.e.,
\be
Q_e=\frac{1}{4\pi}\int *F_{\2}=0\,,\qquad Q_m=\frac{1}{4\pi}\int F_{\2}=0\,.
\ee
This distinguishes the solution from the one presented in \cite{Ma:2026ima}. Furthermore, we can derive the complete thermodynamics of the solution:
\bea
T&=&\lambda _1\frac{ (1-E^2 \mu ^2)^2 \sqrt{\mu ^2-a^2 (1-E^2 \mu ^2)^2} (1-E_0)^2}{4 \pi  \mu
   \big[\mu +\sqrt{\mu ^2-a^2 (1-E^2 \mu ^2)^2}\big]}\,,\cr
\omega_H&=&\lambda _1\frac{ a (1-E^2 \mu ^2)^3 (1-E_0)^2}{2 \mu  \big[\mu +\sqrt{\mu ^2-a^2 (1-E^2 \mu
   ^2)^2}\big]}+E \lambda _2\,,\qquad J=\frac{ \mu a}{(1-E^2 \mu ^2)^2}\,,\cr
   M&=&\lambda _1\frac{ \mu  (1-E_0)^2}{1-E^2 \mu ^2}+2\lambda_2EJ\,,\qquad
S=\frac{2 \pi  \mu  \big[\mu +\sqrt{\mu ^2-a^2 (1-E^2 \mu ^2)^2}\big]}{(1-E^2 \mu ^2)^3}\,.
\eea
The values of \(\lambda_{1,2}\) consistent with both the first law and the CR relation are given by
\bea
\lambda_1=\frac{1}{(1-E_0)^2\sqrt{1-E^2\mu^2}}\,,\qquad \lambda_2=0\,.
\eea
In the limit $E \to 0$, this solution reduces to the Kerr black hole. For \(E\mu<1\), the solution possesses two horizons. The zero of \(\Delta\) defines the black hole horizon \(r_h\), satisfying \(\Delta(r_h)=0\), and the zero of \(I\) defines the cosmological horizon \(r_c\), satisfying \(I(r_c)=0\),
\bea
r_h=\frac{\mu +\sqrt{\mu ^2-a^2 (1-E^2 \mu ^2)^2}}{1-E^2 \mu ^2}\,,\qquad r_c=\frac{1}{E(1-E\mu)}\,.
\eea
A comparison shows that the cosmological horizon always lies outside the black hole horizon
\bea
r_c-r_h=\frac{1+a^2 E^2 (1-E^2 \mu ^2)}{E \Big[1+E \sqrt{\mu ^2-a^2 (1-E^2 \mu
   ^2)^2}\Big]}\geq0\,.
\eea
The static limit of the solution \eqref{E-M field} is given by
\bea
ds^2&=&\frac{\big(1+E^2\mu r(1-x^2)\big)^2}{\Omega^4}\Big[
-\frac{Q}{r^2}dt^2+\frac{r^2}{Q}dr^2+\frac{r^2}{P}\frac{dx^2}{1-x^2}
\Big]+\frac{r^2(1-x^2)P}{\big(1+E^2\mu r(1-x^2)\big)^2}d\varphi^2\,,\cr
A_{\1}&=&\frac{(1-E^2\mu^2x^2)\partial_x\Omega}{ErP}dt\,.
\eea
The corresponding \(\mu=0\) electrovacuum background is precisely the solution obtained in \cite{Ma:2026ima}.

\subsection{KK theory}

The KK theory is a special type of EMD theory \eqref{emd} with $a=\sqrt3$. It arises from the KK circle reduction of $D=5$ pure gravity, via the reduction ansatz
\be
d\hat{s}_5^2=e^{-\frac{1}{\sqrt{3}}\phi}ds_4^2+e^{\frac{2}{\sqrt{3}}\phi}
\big(dz+A_{\1}\big)^2\,.\label{KK reduction}
\ee
The magnetized Kerr black hole in the KK theory from the reduction of ds$_{\rm I,+}^2 + dz^2$, with $\varphi\rightarrow\varphi + \ft12 B z$, was considered in \cite{Ma:2026rro}. In this subsection, we replace the seed solution $ds_{\rm I,+}^2$ by $ds_{\rm II,+}^2$. The resulting metric is
\bea
ds^2&=&\frac{B^4 \big[P (r^2-a^2)^2 (1-x^2)-a^2 Q (1+x^2)^2\big]}{16 \sqrt{H} \mathbb{L} \Omega ^4}\Big(d\varphi +\frac{a \mathbb{M} dt}{P (r^2-a^2)^2 (1-x^2)-a^2 Q (1+x^2)^2}\Big)^2\cr
&&-dt^2\frac{B^2 P Q (1-x^2)}{16 \sqrt{H}}\Big(\frac{B^2 \mathbb{L}}{P (r^2-a^2)^2 (1-x^2)-a^2 Q (1+x^2)^2}+\frac{4 (1-B_0)^2}{\Omega ^4}\Big)\cr
&&+\frac{B^4 \sqrt{H} \mathbb{L}}{16}  \Big(\frac{dr^2}{Q}+\frac{1}{P}\frac{dx^2}{1-x^2}\Big)\,,\cr
A_{\1}&=&\frac{B(1-B_0)}{2H\mathbb{L}\Omega^4}\Big[
a\mathbb{M}dt+\big[P (r^2-a^2)^2 (1-x^2)-a^2 Q (1+x^2)^2\big]d\varphi
\Big]\,,\qquad \phi=\frac{\sqrt{3}}{2}\log H\,,\cr
H&=&1+\frac{4(1-B_0)^2}{B^2\mathbb{L}\Omega^4}\big[P (r^2-a^2)^2 (1-x^2)-a^2 Q (1+x^2)^2\big]\,.
\eea
Our global analysis indicates that we need to perform the following coordinate transformations
\bea
&&\varphi\quad \rightarrow\quad \varphi'=\frac{\varphi}{P_{0}}\,,\qquad t\quad \rightarrow \quad t'=t+\alpha\frac{\varphi}{P_{0}}\,,\cr
&&\alpha=\frac{aB^4}{8(1-B_0)^4}\,,\qquad P_0=\frac{B^4(1+B^2\mu^2)}{16(1-B_0)^2}\,,
\eea
so that the resulting $\varphi$ makes the Killing vector degenerate at $x=\pm 1$, and
the absence of the conic singularity requires it to have period $2\pi$. It can be verified, as in the case of G-KBR black hole, the above solution carries no electric or magnetic charges; therefore, it can be genuinely viewed as a Kerr black hole immersed in a uniform external magnetic field. Unfortunately, the metric has naked CTCs, as $g_{\varphi\varphi}$ can be negative outside the horizon. To see this specifically, if we fix $x=\pm 1$ and take $r\rightarrow$ to infinity, we have
\be
g_{\varphi\varphi}(r\rightarrow\infty)=-\frac{a(1-B_0)}{\sqrt{2}B(1+B^2\mu^2)^{3/2}}
\frac{1}{\sqrt{1-x^2}}-\frac{11\mu a}{2\sqrt{2}(1+B^2\mu^2)^2}\,.
\ee
On the other hand, $g_{\varphi\varphi}=0$ for any given $r$ at $x=\pm 1$. These two facts imply that there must be $g_{\varphi\varphi}<0$ regions in the vicinity of both the north and south poles at large $r$.

\section{Conclusions}
\label{sec:conclusion}

In this paper, we presented a new family of Ricci-flat rotating metrics, referred to as Solution II, by demagnetizing the G-KBR black hole. Together with Solution I, previously obtained by demagnetizing the KBR black hole, we now have two families of Ricci-flat Kerr deformations, both of which are characterized by three parameters \((\mu,a,B)\).
Both families have cohomogeneity two, depending on two coordinates $(r,x=\cos\theta)$, in addition to the two Killing coordinates $(t,\varphi)$; they both admit two branches associated with the $\pm$ signs of \(\Omega\), when written in $(r,x)$ coordinates, with $g_{rx}=0$. For nonzero \(B\), the two branches within each family can be unified in the \((r,\Omega)\) coordinates, but with non-vanishing $g_{r\Omega}$, up to a trombone rescaling. When \(B=0\) and hence $\Omega^2=1$, however, the coordinate transformation becomes singular, so this unified description degenerates and the two branches become distinct. In the \(B\to0\) limit, one of the two branches necessarily becomes divergent. A suitable trombone rescaling is therefore required to ensure that both branches admit smooth \(B\to0\) limits. We then find that the two limiting geometries are embedded in the Minkowski and Levi-Civita backgrounds, respectively. We denote the former by the plus branch and the latter by the minus branch. This implies that the Minkowski vacuum and the Levi-Civita background, which are drastically different, become unified under the \(B\neq0\) deformation, and this unification persists after the black holes are introduced. We further find that these two \(B\)-deformed vacua correspond precisely to two disconnected sectors of the same metric in the Kundt class.

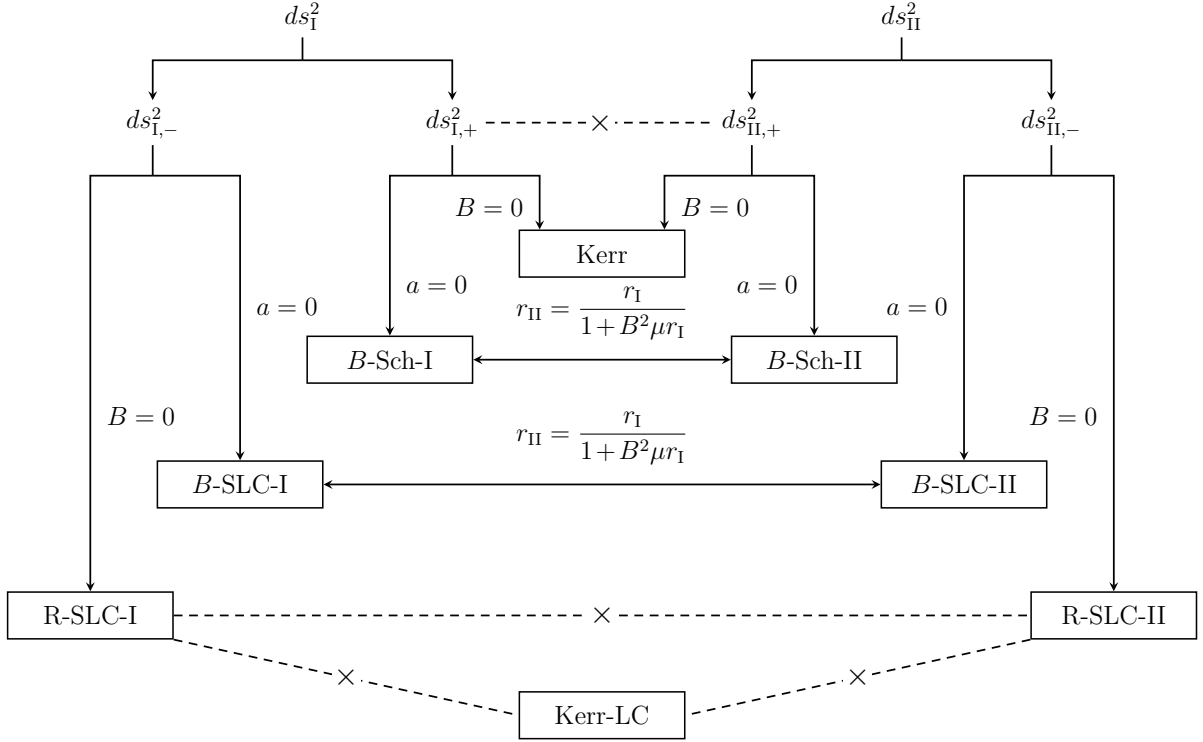
\begin{figure}[htbp]
\centering
\resizebox{\textwidth}{!}{%
\begin{tikzpicture}[
  font=\large,
  >=stealth,
  every node/.style={inner sep=3pt},
  metric/.style={draw, line width=0.7pt,
    minimum width=2.65cm, minimum height=0.75cm},
  limit/.style={->, line width=0.8pt},
  equivalent/.style={<->, line width=0.8pt},
  inequivalent/.style={-, line width=0.8pt,
    dash pattern=on 4pt off 3pt},
  cross/.style={fill=white, inner sep=1pt, font=\Large}
]

\node (I)   at (-4.8, 0)    {\(ds_{\mathrm{I}}^{2}\)};
\node (II)  at ( 4.8, 0)    {\(ds_{\mathrm{II}}^{2}\)};
\node (Im)  at (-7.2,-1.7)  {\(ds_{\mathrm{I},-}^{2}\)};
\node (Ip)  at (-2.4,-1.7)  {\(ds_{\mathrm{I},+}^{2}\)};
\node (IIp) at ( 2.4,-1.7)  {\(ds_{\mathrm{II},+}^{2}\)};
\node (IIm) at ( 7.2,-1.7)  {\(ds_{\mathrm{II},-}^{2}\)};

\node[metric] (Kerr)  at ( 0,  -3.8)  {Kerr};
\node[metric] (BSI)   at (-3.4,-5.5)  {\(B\)-Sch-I};
\node[metric] (BSII)  at ( 3.4,-5.5)  {\(B\)-Sch-II};
\node[metric] (BLCI)  at (-5.8,-7.5)  {\(B\)-SLC-I};
\node[metric] (BLCII) at ( 5.8,-7.5)  {\(B\)-SLC-II};
\node[metric] (RI)    at (-8.2,-9.6)  {R-SLC-I};
\node[metric] (RII)   at ( 8.2,-9.6)  {R-SLC-II};
\node[metric] (KLC)   at ( 0, -11.2)  {Kerr-LC};

\coordinate (sI)  at (-4.8,-0.7);
\coordinate (sII) at ( 4.8,-0.7);
\draw[line width=0.8pt] (I.south) -- (sI);
\draw[line width=0.8pt] (II.south) -- (sII);
\draw[limit] (sI)  -| (Im.north);
\draw[limit] (sI)  -| (Ip.north);
\draw[limit] (sII) -| (IIp.north);
\draw[limit] (sII) -| (IIm.north);

\coordinate (sIm)  at (-7.2,-2.55);
\coordinate (sIp)  at (-2.4,-2.55);
\coordinate (sIIp) at ( 2.4,-2.55);
\coordinate (sIIm) at ( 7.2,-2.55);
\draw[line width=0.8pt] (Im.south)  -- (sIm);
\draw[line width=0.8pt] (Ip.south)  -- (sIp);
\draw[line width=0.8pt] (IIp.south) -- (sIIp);
\draw[line width=0.8pt] (IIm.south) -- (sIIm);

\draw[limit] (sIp)  -- (-1.0,-2.55) -- ([xshift=-1cm]Kerr.north);
\draw[limit] (sIIp) -- ( 1.0,-2.55) -- ([xshift= 1cm]Kerr.north);
\node[anchor=east] at (-1.15,-3.05) {\(B=0\)};
\node[anchor=west] at ( 1.15,-3.05) {\(B=0\)};

\draw[limit] (sIp)  -| (BSI.north);
\draw[limit] (sIIp) -| (BSII.north);
\node[anchor=west] at (-3.25,-4.3) {\(a=0\)};
\node[anchor=east] at ( 3.25,-4.3) {\(a=0\)};

\draw[limit] (sIm)  -| (BLCI.north);
\draw[limit] (sIIm) -| (BLCII.north);
\draw[limit] (sIm)  -| (RI.north);
\draw[limit] (sIIm) -| (RII.north);
\node[anchor=west] at (-5.65,-4.65) {\(a=0\)};
\node[anchor=east] at ( 5.65,-4.65) {\(a=0\)};
\node[anchor=west] at (-8.05,-6.4)  {\(B=0\)};
\node[anchor=east] at ( 8.05,-6.4)  {\(B=0\)};

\draw[equivalent] (BSI.east) --
  node[midway,above=6pt]
    {\(\displaystyle r_{\mathrm{II}}
      =\frac{r_{\mathrm{I}}}{1+B^{2}\mu r_{\mathrm{I}}}\)}
  (BSII.west);
\draw[equivalent] (BLCI.east) --
  node[midway,above=6pt]
    {\(\displaystyle r_{\mathrm{II}}
      =\frac{r_{\mathrm{I}}}{1+B^{2}\mu r_{\mathrm{I}}}\)}
  (BLCII.west);

\draw[inequivalent] (Ip.east) --
  node[cross] {\(\times\)} (IIp.west);
\draw[inequivalent] (RI.east) --
  node[cross] {\(\times\)} (RII.west);
\draw[inequivalent] (RI.south east) --
  node[cross] {\(\times\)} (KLC.west);
\draw[inequivalent] (RII.south west) --
  node[cross] {\(\times\)} (KLC.east);

\end{tikzpicture}%
}
\caption{\small This figure illustrates various limits of Solution I and II and their connections. The dashed line with a cross means that the two connected metrics are locally inequivalent.}\label{figure}
\end{figure}

In the static case \(a=0\), the two branches, SLC-I and SLC-II, are locally equivalent and can be mapped into each other by a simple coordinate transformation. It should be emphasized, however, that even with the $B$-deformation present, \(x=\pm1\) remain curvature singularities of both Levi-Civita branches. However, we find that introducing a rotation can potentially resolve the singularity. For \(a\neq0\) and \(B=0\), the minus branches provide two rotating generalizations of the SLC metric, which we denote by R-SLC-I and R-SLC-II, respectively. Together with the KLC spacetime in the literature, we now have three different ways of introducing rotations into the SLC metric. We have presented strong evidence that these three metrics are all locally inequivalent. In Fig.~\ref{figure}, we illustrate the two families, their various special limits, and the corresponding connections.

The global analysis of the metric $ds_{\rm I,+}^2$ was presented in \cite{Ma:2026otg}. In this paper, we studied the global structures of $ds_{\rm I,-}^2$ and $ds_{\rm II,\pm}^2$. We focused on event horizons, naked CTCs and ergoregions. We derived the first law of black hole thermodynamics. We found that $ds_{\rm I,-}^2$ describes a black hole with no naked CTCs and a finite ergoregion. The solution $ds_{\rm II,+}^2$ may describe a black hole with no naked CTCs and disjoint ergoregions that could extend to infinity. In both cases, the first law takes the same form. The solution $ds_{\rm II,-}^2$ is very unusual, describing a black hole time machine.

Our result establishes that all the known black holes in external magnetic fields in Einstein-Maxwell theory have Ricci-flat origin via Ernst transformation, and the new Ricci-flat seeds provide new possibilities of magnetization. We studied the remagnetization of our Ricci-flat seeds in Einstein-Maxwell and KK theories. The general magnetized solutions in both theories involve four parameters, $(B,\mu, a, {\cal B})$. Kerr-Melvin solution with magnetic parameter ${\cal B}$ arises if we take $B=0$. In order to reproduce the KBR and G-KBR solutions, the parameter ${\cal B}$ must be chosen to satisfy a specific relation with $B$, so that the distinction of the two branches disappear. Furthermore, the Ernst transformation may require a specific trombone rescaling with the conformal factor determined by $B$. This suggests that both KBR and G-KBR are fine-tuned results of a more general class of solutions. Applying the magnetization procedure after the continuation \(B=\mathrm{i}E\) yields a charge-free rotating Einstein-Maxwell solution with a cosmological horizon, without introducing a cosmological constant. The KK construction also produces an uncharged rotating solution immersed in an external magnetic field, although we find it gives rise to naked CTCs. A better understanding of achieving charge-free magnetization and simultaneously avoiding naked CTCs is needed and is left for future investigation.

{\bf Note Added}: With the help of AI, the equations \eqref{M1L1}, \eqref{M2L2B0} and \eqref{LH} are significantly simplified compared to the old version.

\section*{Acknowledgement}

L.M.~is supported in part by National Natural Science Foundation of China (NSFC) grant No.~12447138, Postdoctoral Fellowship Program of CPSF Grant No.~GZC20241211, the China Postdoctoral Science Foundation under Grant No.~2024M762338 and the National Key Research and Development Program No.~2022YFE0134300. H.L.~is supported in part by the NSFC grants No.~12375052 and No.~11935009. The work is also supported in part by the Tianjin University Self-Innovation Fund Extreme Basic Research Project Grant No.~2025XJ21-0007.

\end{document}